# Filling in the gaps: The nature of light transmission through solvent-filled inverse opal photonic crystals

Alex Lonergan[1], Changyu Hu[1], and Colm O'Dwyer[1,2,3,4]*

[1]School of Chemistry, University College Cork, Cork, T12 YN60, Ireland
[2]Micro-Nano Systems Centre, Tyndall National Institute, Lee Maltings, Cork, T12 R5CP, Ireland
[3]AMBER@CRANN, Trinity College Dublin, Dublin 2, Ireland
[4]Environmental Research Institute, University College Cork, Lee Road, Cork T23 XE10, Ireland

## Abstract

Understanding the nature of light transmission and the photonic bandgap in inverse opal photonic crystals is essential for linking their optical characteristics to any application. This is especially important when these structures are examined in liquids or solvents. Knowledge of the true correlation between the nature of the inverse opal (IO) photonic bandgap, their structure, and the theories that describe their optical spectra is surprisingly limited compared to colloidal opals or more classical photonic crystal structures. We examined $TiO_2$ and $SnO_2$ IOs in a range of common solvents to solve the conflict between Bragg-Snell theory, optical and physical measurements by a comprehensive angle-resolved light transmission study coupled to microscopy examination of the IO structure. Tuning the position of the photonic bandgap and index contrast by solvent infiltration of each inverse opal requires a modification to the Bragg-Snell theory and the photonic crystal unit cell definition. We also demonstrate experimentally and theroetically that low fill factors are caused by less desne material infilling *all* interstitial vancancies in the opal template to form an IO. By also including an optical interference condition for inverse opals with an effective refractive index greater than its substrate, and an alternative internal refraction angle in the substrate, angle-resolved transmission spectra for inverse opals are now consistent with physical measurements. This work now allows an accurate correlation between the true response of an IO to the index contrast with a solvent, how an IO is infilled, and the directionality and bandwidth of the photonic bandgap. As control in functional photonic materials becomes more prevalent outside of optics and photonics, such as biosensing and energy storage, for example, a comprehensive and consistent correlation between photonic crystals structures and their primary optical signatures is a fundamental requirement for application.

# Introduction

Photonic crystal structures can mold the flow of light because their structure results in a photonic bandgap where photons are 'forbidden' at certain frequencies. It is an effect that natural evolution has given to butterfly wings, peacock feathers and beetle shells that show iridescent colors from diffraction, not dyes. Ordered, porous materials with a repeating nanostructured lattice can interfere with light incident on the surface giving rise to what we often refer to as structural color. Since their conception[1,2], the prospect of an alternating dielectric contrast in photonic crystal materials has made them attractive candidates for a wide variety of practical applications[3]. Analogous to the electronic bandgap in semiconductor devices, photonic crystal materials induce a photonic bandgap (PBG) [4], whereby certain frequency ranges are prohibited as a result of structural interference with light. The efficacy of a PBG in the depletion of a frequency range is dependent on the magnitude of the dielectric contrast in the material. It is thought that a full PBG is possible for a material featuring a relatively high refractive index $n \geq 2.8$[5], compared to an air background. Partial depletion of a frequency range is also possible, creating a pseudo-photonic bandgap (p-PBG), whereby an insufficiently high refractive index contrast is present in the material[6].

One of the most appealing aspects of photonic crystal structures lies in the ability to manufacture these structures from a very wide range of materials, from changes in composition to different material types (solids, polymers etc.). Colloidal crystals assembled using polystyrene (PS), poly(methyl methacrylate) (PMMA) or silica spheres, adopt a photonic crystal structure, typically reporting a p-PBG[7,8]. However, this ordered sphere structure can also act as a template for the creation of an inverse opal (IO) material[9-12]. Sphere templates can be infilled with liquid precursors creating a periodic repeating structure or photonic crystal, notionally considered to be the inverse of the colloidal opal template. Using a suitable precursor, a plethora of inverse opal materials can be formed this way [13] and by the rational choice of refractive index contrast[14], the position of the PBG can be tuned.

Due to this inherent flexibility in material choice and fabrication process, inverse opal materials with a tuneable PBG[15] have garnered great interest. The PBG can be tuned via control over the refractive index of the material, the dimensions of the repeating structure and the refractive index of the background material. An adjustable PBG creates a system with high sensitivity to minor changes in material properties through monitoring of the modified PBG[16]. As such, inverse opal materials are incorporated into a wide variety of disciplines, including optical[17-19], energy storage[20,21], biological[22-24] and medical fields[25,26]. The novel, repeating structure can act as an optical waveguide[27], a refractive index sensor[28,29], a biosensor for virus detection[25,30], an enhanced material for solar absorption[31,32], the gain medium in lasing materials[33,34] and structured electrode materials in Li-ion batteries[35-39].

The optical properties of photonic crystal structures have been thoroughly studied in the case of opal colloidal crystals[40,41]. The physics of opal photonic crystals is mature, and the nature of the PBG is well defined with respect to the periodic structure, dispersion relations and how such structures control the propagation of electromagnetic radiation. The diameters of opal microstructures and their refractive indices have been repeatedly and successfully assessed to a high degree of certainty by inspection of the optical transmission spectrum of the structure[42,43]. The optical properties of a wide range of inverse opal materials have also been studied extensively[44-47]. Unfortunately, the spectra of inverse opal materials are often not as easily interpreted as their opal counterparts, and we note a dearth of literature for inverse opal optics compared to colloidal photonic crystals. In some cases[48,49], particularly those featuring alternative fabrication methods such as atomic layer deposition, estimations of material parameters using the PBG report good agreement with microscopy. Inverse opals prepared via atomic layer deposition are notably different from those obtained via calcination of a sacrificial opal template, with some inverse opals formed by atomic layer deposition better described as hollow shells of the deposited material, with an opal structure as opposed to a connected scaffold of interstitial material. In other cases[50,51], such as those with inverse opals prepared via annealing an opal template, there is a clear difference between the estimation of material properties from the spectral data and the independently measured microscopy data. It has been suggested that the optical behavior of inverse opal materials is inherently different from the reciprocal of an opal, with the possibility of an axial compression[51], reduced unit cell parameter[52] and, most commonly, reduced crystalline material volume fractions[53-57] suggested as possible explanations. Currently, there is an absence of consensus on the explanation underpinning the differences observed in the PBG for inverse opal structures. This important issue underpins every investigation using inverse opals where their spectral response is required.

We present a detailed investigation into the nature of the PBG and angle-resolved structural color of $TiO_2$ and $SnO_2$ inverse opal structures, focusing on their spectral behavior in air, but also when infilled with a range of solvents. Only a surprisingly small number of publications[51,58-62] have addressed the interesting optical properties of inverse opal materials in solvent media, including the nature of PBG bandwidth, directionality, and sensitivity of the degree of infilling. We present a full study on the effects of solvent infiltration in inverse opal materials and, to our knowledge, the first study to systematically investigate the effects of solvent infilling with variable incident angle that is fully consistent with the actual IO structure. The behavior of the PBG in this scenario has only been speculated to date[63], with an assumed, though not tested, analogous response to the IO in air. To provide a complete assessment of the IO photonic crystals, the optical data will also be evaluated in air and in solvents at normal incidence. By monitoring changes to the optical spectrum at normal incidence we suggest the most accurate model for approximation of the effective refractive index of inverse opal materials and verify the influence of fill factor by computational modeling of the PBG and the IO structure itself. The fill factor definition is important, as it is usually defined in relation to the free space than can be filled; a variation assumes some of that space is not filled. We also test this aspect but also consider a situation where material exists in all available voids in the opal, but where the degree of interstitial filling itself is partially porous.

We also put forward an argument based on thin-film interference to explain the differences between the spectral response of the PBG of IO materials in air compared to those infilled with solvents on transparent substrates required for light transmission. Our analysis suggests that the Bragg-Snell model describing the PBG behavior in air is not directly applicable to the solvent-infilled IO case, and a modification to the internal angle is required to fit the data. We posit that this is due to the effective refractive index of the photonic crystal layer exceeding the substrate value, introducing a change to the constructive interference condition.

## Materials and methods

**Substrate preparation and treatment**

Polystyrene spheres of 500 and 350 nm diameter in a 2.5 wt.% aqueous suspension were purchased from Polysciences Inc. The spheres contained some anionic charge from the sulfate ester used in formation, resulting in a net negative charge for the suspension. All sphere suspensions were used as received. Fluorine-doped tin oxide (FTO)-coated soda-lime glass was purchased from Solaronix SA and used as a transparent conductive substrate for colloidal crystal templates. Substrate glass of thickness 2.2 mm was cut into pieces of sizes 10 × 25 mm. All substrates were cleaned via successive sonication in acetone (reagent grade 99.5%; Sigma Aldrich), isopropyl alcohol (reagent grade 99.5%; Sigma Aldrich) and deionized water. Following cleaning, all samples were dried in an inert atmosphere at room temperature. Surface coating areas of 10 × 10 mm were prepared on the conductive surface of the substrate by covering the remainder of the exposed surface with Kapton tape. Kapton tape was applied to the front and back of the substrate to ensure the formation of the colloidal crystal template on the desired surface area. Immediately prior to photonic crystal formation, substrates were treated in a Novascan PSD Pro Series digital UV-Ozone system for 1 h to ensure surface hydrophilicity of the substrate.

**Polystyrene opal template formation**

Directly following UV-Ozone treatment, treated glass substrates were slowly immersed into pre-heated vials of 2.5 wt.% polystyrene sphere suspensions at a rate of 1 mm min$^{-1}$ using a MTI PTL-MM01 Dip Coater. Sphere suspensions were gently heated to temperatures of ~50 °C prior to substrate insertion, showing a close-packed and uniformly thick opal coating[64]. Substrates were also inclined at a slight angle of ~10° - 20° from the vertical, so as to improve adhesion to the conductive layer of the substrate[65]. Substrate areas immersed in sphere suspensions were held still for 10 min to accommodate for a period allowing the sphere suspension to settle to a minimum energy state from surfactant-mediated repulsion. Substrates were then withdrawn at a rate of 1 mm min$^{-1}$ and allowed to dry in an inert atmosphere at room temperature.

**$TiO_2$ and $SnO_2$ inverse opal formation**

$TiO_2$ and $SnO_2$ inverse opals were formed via infiltration of a polystyrene sphere template with a suitable precursor with subsequent calcination of the material in air at 450 °C for 1 hr. $TiO_2$ inverse opals were formed using a 0.1 M solution of $TiCl_4$ in isopropyl alcohol. A titanium (IV) chloride tetrahydrofuran complex ($TiCl_4 \cdot 2THF$, 97%; Sigma Aldrich) was used as the source of $TiCl_4$. Likewise, a 0.1 M solution of $SnCl_2$ was used in the formation of the $SnO_2$ inverse opals, with tin(II) chloride dihydrate (98%; Arcos Organics) used as the source of $SnCl_2$.

**Microscopy and optical characterization**

A Zeiss Supra 40 high-resolution SEM at an accelerating voltage of 10 kV was used to carry out all relevant SEM imaging. Analysis of SEM images and feature dimensions was performed using ImageJ software. Raman scattering analysis was carried out using a Renishaw InVia Raman Spectrometer in conjunction with a 30 mW Ar$^+$ laser featuring an excitation wavelength of 514 nm. The laser was focused using a 40× objective lens and collected using a RenCam CCD camera. Optical transmission analysis was carried out using a tungsten-halogen lamp with an operating wavelength range 400 – 2200 nm, purchased from Thorlabs Inc. and a UV-Visible spectrometer (USB2000 + VIS-NIR-ES) with an operational range of 350 – 1000 nm, from Ocean Optics Inc. The angle of incidence was varied for transmission measurements using a motorized rotational stage (ELL8; Thor Labs Inc.).

**Photonic band structure modeling**

The inverse opal (IO) photonic crystal structure was analyzed using the plane wave expansion (PWE) method by using the BandSOLVE software package (Rsoft Design Group, Inc.). A three-dimensional face-centered cubic computational model was built based on SEM image observations. The parameters in the simulation model are consistent with theoretical calculations. The effects of filling factor and positions of each layers' lattice were designed to specific configurations of IO photonic crystal structures. A perfect match layer is located around the whole structure as a perfect absorber boundary condition. Hybrid polarizations of full and partial band diagrams were respectively presented with the corresponding experiment comparisons.

## Results and Discussion

### Inverse opal photonic crystals of $TiO_2$ and $SnO_2$

Inverse opal (IO) photonic crystal materials of $TiO_2$ and $SnO_2$, were first fully characterized in air to accurately define their structure, dimensions, and spectral response. A detailed study of the IO response in air provides a useful and necessary foundation for accurate interpretation of light transmission and propagation in the IO when infilled with solvents. Using a 500 nm polystyrene opal template, IOs of $TiO_2$ and $SnO_2$ were fabricated using a simple sol-gel infiltration of the metal precursor into the highly ordered colloidal crystal template. Optical images of the IOs prepared on fluorine-doped tin oxide (FTO) coated glass substrates can be viewed for $TiO_2$ and $SnO_2$ in Figs 1(a) and (c), respectively. The highly ordered, repeating structure of the IO confirms the expected (111) planar geometry retained from the colloidal crystal template, as anticipated from a high quality IO. Figures 1 (b) and (d) respectively, correspond to a typical SEM image of the IO surface for prepared $TiO_2$ and $SnO_2$ IO where our fabrication method ensures consistently similar and fully formed three dimensional IO architectures throughout. From SEM images, typical $TiO_2$ and $SnO_2$ IOs feature ~10 layers of material, giving approximate film thicknesses of 3 - 4 μm (see Supporting Information Figs S1 & S2).

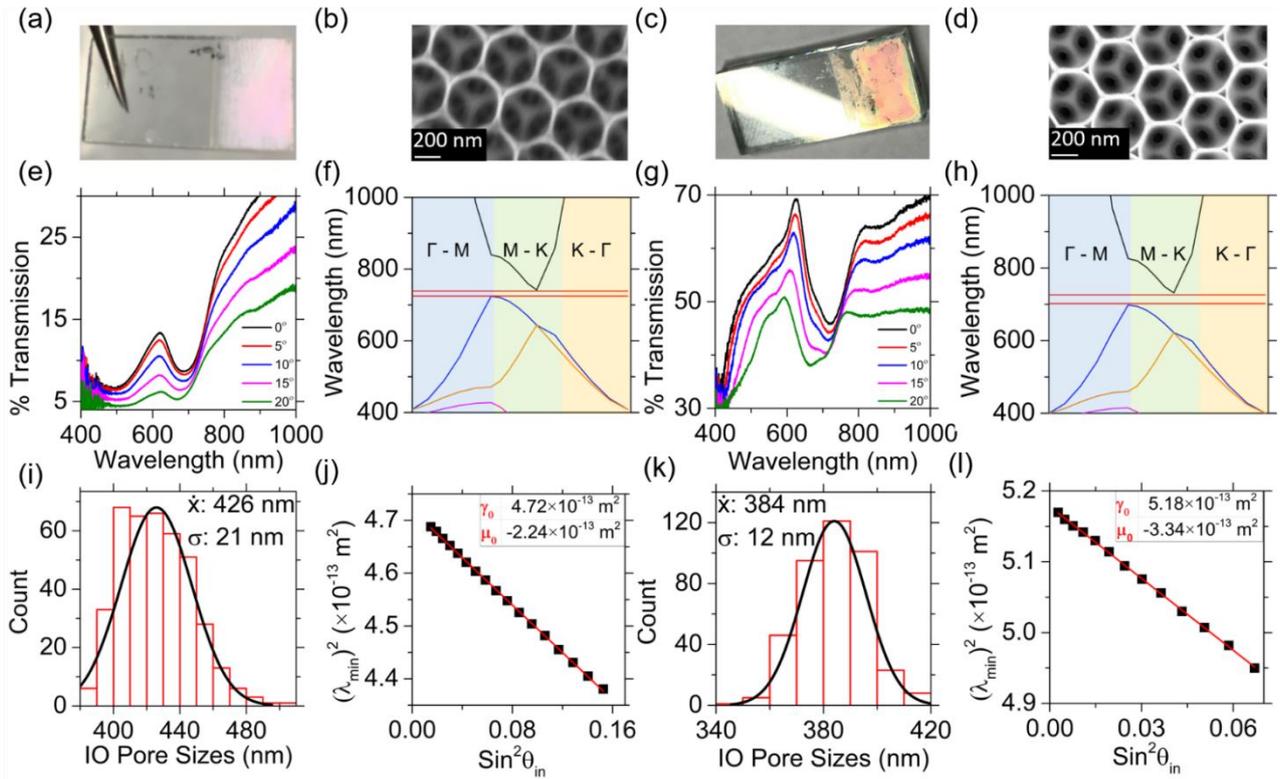

**Figure 1. Defining the structure and PBG of TiO$_2$ and SnO$_2$ IOs.** The (a,c) picture and (b,d) SEM image confirm very high-quality TiO$_2$ and SnO$_2$ IOs respectively. (e,g) Transmission spectra for both IOs using visible light incident at 0°, 5°, 10°, 15°, and 20° are corroborated by (f,h) band structure diagrams from plane wave expansion with a highlighted photonic bandgap region along Γ– M, M – K, and K – Γ directions using fill factors of $\varphi_{IO}$ = 0.26 and $\varphi_{IO}$ = 0.18 (the latter indicates porous material within *all* voids forming the IO). The region of inhibited frequency propagation is highlighted in red, with the center of this region being taken as an estimate for the position of the transmission minimum ($\lambda_{min}$). Size distribution analysis for center-to-center pore distances (See Supporting Information Figs. S1 and S2 for measurement details) for (I,k) TiO$_2$ and SnO$_2$ IO respectively. (j,l) Bragg-Snell plots of $\lambda_{min}^2$ vs Sin$^2$θ showing slope ($\mu_0$) and intercept ($\gamma_0$) per Eq. (2) for both IOs.

Gauging periodicity and consistent with an IO is crucial when the diffraction effect and associated theory are based on order. In our case, we obtained normal distributions of the centre-to-centre pore distances for TiO$_2$ and SnO$_2$ IOs, as seen in Figs. 1 (i) and (k) from a large data set of measurement from SEM. For TiO$_2$ IOs, a sample mean of centre-to-centre pore size of 426 nm (σ = 21 nm) was found. By comparison, SnO$_2$ IOs exhibit a corresponding mean centre-to-centre pore distance of 384 nm (σ = 12 nm). Compared to the sacrificial colloidal crystal of 500 nm diameter used as a template, there is a measurable shrinkage (15% for TiO$_2$, 23 % for SnO$_2$) upon inversion and formation of the inverse opal structure, but no anisotropic changes to pore wall diameter or shape are obvious. There are previous reports on annleaing or calcination-based approaches to IOs where a similar reduction from sphere diameter to IO macropore center-to-center distance in the material periodicity[9,66-68]. Raman scattering analysis of IO structures show characteristic phonon modes for SnO$_2$ and anatase phase TiO$_2$ (see Supporting Information, Fig. S3). Band structure diagrams scaled to the wavelength range of the optical transmission spectra are shown for TiO$_2$ (D = 400 nm) and SnO$_2$ (D = 384 nm) IOs in air in Figs. 1 (f) and (h) respectively. The fill fraction of crystalline material, $\varphi_{IO}$, used in simulations was varied to match experimental findings presented in Fig. 2. Previous work by TEM confirmed the octahedral and tetrahedral vacancies of the opal template are infilled with material, but this filling is porous and represented by the reduced fill factor in our analysis. Simulations were performed for a TiO$_2$ IO with $\varphi_{IO} = 0.18$ and a SnO$_2$ with $\varphi_{IO} = 0.26$ by introducing porosity in the infilling of all tetrahedral and octahedral vacancies of the parent opal template. A full photonic bandgap across all directions is present for both IO structures in air. For the TiO$_2$ IO, the predicted PBG extends from 724 – 739 nm, giving an approximate value for $\lambda_{min}$ of 731 nm. In the case of the SnO$_2$ IO, the projected PBG spans a wavelength range of 702 – 728 nm, yielding an estimated value for $\lambda_{min}$ of 715 nm. Further analysis of the band structures for TiO$_2$ and SnO$_2$ IOs can be found in the Supporting Information in Fig. S11.

Optical transmission spectra at various angles of incidence for a typical TiO$_2$ and SnO$_2$ IO are shown in Figs. 1 (e) and (g), respectively. At 0° incidence, distinct transmission minima are present at ~690 nm for TiO$_2$ and ~720 nm for SnO$_2$. Comparing these experimentally determined $\lambda_{min}$ positions with the simulated estimates from the band structures in Figs. 1 (f) and (h), we find an excellent agreement for the SnO$_2$ IO transmission minimum. For the SnO$_2$ IO, there is ~5 nm difference between the simulated result of 715 nm and the experimentally determined 720 nm for the transmission minimum. Conversely, the TiO$_2$ IO exhibits a much larger difference of 41 nm between the simulated and experimental $\lambda_{min}$ value. It would seem that the band structure diagram more accurately predicts the position of $\lambda_{min}$ for structures with $\varphi_{IO} = 0.26$, a discrepancy that is unique to the IO in air, as will be shown.

Looking at Figs. 1 (e) and (g), at higher angles of incidence we see a blue-shift of the transmission minima, just as predicted by Bragg-Snell analysis. For a crystal plane (hkl) with interplanar spacing $d_{hkl}$, an IO material with an effective refractive index $n_{eff}$, some solvent of refractive index $n_{sol}$ filling the pores, a Bragg resonance order $m$ and some angle $\theta$ between the incident light and the normal to the crystal plane surface, the transmission minimum $\lambda_{hkl}$ can be found via the Bragg-Snell model as follows:

$$\lambda_{hkl} = \frac{2d_{hkl}}{m}\sqrt{n_{eff}^2 - n_{sol}^2\sin^2\theta} \qquad (1)$$

Assuming the first-order resonance with $m$ = 1, squaring both sides of Eq. 1 for an FCC (111) plane yields:

$$\lambda_{min}^2 = 4d_{111}^2 n_{eff}^2 - 4d_{111}^2 n_{sol}^2\sin^2\theta \qquad (2)$$

With respect to Eq. (2), a plot of the observed minimum wavelength squared versus the sine of the angle of incidence squared should yield a linearly regressive plot with a slope $\mu_0 = -4d_{111}^2 n_{sol}^2$ and an intercept $\gamma_0 = 4d_{111}^2 n_{eff}^2$. From this, the interplanar spacing $d_{111}$ and hence the center-to-center pore distance $D$ can be found via the slope. The effective refractive index $n_{eff}$ of the IO material can be estimated by dividing the intercept by the slope. Evaluating $n_{eff}$ via this method does not require a known value for $d_{111}$, allowing a value to be found independent of the spacing determined from the spectra. As previously reported for TiO$_2$ IOs[52], applying a reduced interplanar spacing of $d = \sqrt{\frac{1}{3}} D$ as opposed to the standard interplanar spacing of $d = \sqrt{\frac{2}{3}} D$ appears to better model the system, in terms of estimating a value for the centre-to-centre pore distance.

Figure 1 (j) shows the characteristic optical transmission data for a typical TiO$_2$ IO as detailed in Eq. (2). Applying the standard interplanar spacing $d = \sqrt{\frac{2}{3}} D$ to the slope data, a center-to-center pore distance of ~290 nm is found, drastically different from the estimate of ~426 nm using the mean size from measurements. Applying the reduced interplanar spacing ($d = \sqrt{\frac{1}{3}} D$) model provides an estimate of approximately 410 nm, more in line with expectations from SEM data. Conversely, the opposite case proves true for the SnO$_2$ IO. From Bragg-Snell data in Fig. 1 (l) for a typical SnO$_2$ IO, the center-to-center pore distance is estimated to be ~354 nm. This is similar to the SEM measured mean of ~384 nm for SnO$_2$ IOs. A reduced interplanar spacing applied to the SnO$_2$ data greatly overestimates the center-to-center pore distances (~500 nm). As we will see later, these discrepancies have resulted in fill factors for IOs reported in the literature that are not consistent with the reported structure.

For the TiO$_2$ and SnO$_2$ IOs in air, $n_{eff}$ can also be estimated from Figs. 1 (j) and (l) to give approximate values of 1.45 and 1.25 respectively. Currently, there are several methods used in a wide variety of literature to approximate the effective refractive index of composite material[69,70]. Two common models applied to inverse opal materials to determine $n_{eff}$ include the Drude[16,48,52,71-73] and Parallel[58-62] models. The Drude model for inverse opal systems can be formalized as:

$$n_{eff} = \sqrt{n_{IO}^2 \varphi_{IO} + n_{sol}^2 \varphi_{sol}} \qquad (3)$$

The Parallel model for an inverse opal system can be written as:

$$n_{eff} = n_{IO}\varphi_{IO} + n_{sol}\varphi_{sol} \qquad (4)$$

For an inverse opal system, $n_{IO}$ refers to the refractive index of the crystalline IO material, $n_{sol}$ is the refractive index of the surrounding medium, $\varphi_{IO}$ constitutes the volume filling fraction of IO material in the system and $\varphi_{sol}$ is the volume filling fraction of the surrounding medium. Naturally, the sum of both volume filling fractions should be unity. For an IO structure, an optimally filled structure should feature $\varphi_{IO}$ = 0.26. Admittedly, for photonic crystal systems featuring materials with moderate refractive index values, the differences between these two models are minimal. A significant difference is observed for materials of high refractive index. Applying Eqs. (3) and (4) to an anatase phase $TiO_2$ IO with $n_{IO}$ = 2.49[74], values for $n_{eff}$ can be estimated as 1.39 and 1.53, respectively. From the optical transmission data, the calculated $n_{eff}$ value of 1.45 lies between these theoretical estimates. In the case of the $SnO_2$ IO with $n_{IO}$ = 2.01[50,57], $n_{eff}$ can be estimated as 1.26 and 1.34, through use of Eqs. (3) and (4) respectively. The experimental value for $n_{eff}$ determined from the analysis of the optical transmission spectrum yields a value of 1.25, much closer to the value determined via the Parallel model.

**Solvent filling of inverse opals and the nature of structural color**

Monitoring the spectral change in the transmission minima is useful for determining how accurately a particular refractive index model can predict the true effective refractive index. For inverse opals filled with solvents, the effect of reduced fill factor $\varphi_{IO}$ < 0.26, the directionality of the PBG and the magnitude of reflectivity of forbidden wavelengths, is surprisingly less well known. Figures 2 (c) and (d) show the evolution of the Bragg-Snell transmission minimum at normal incidence ($\theta$ = 0°), with the addition of solvents into the porous crystalline network of an IO, for a typical $TiO_2$ and $SnO_2$ IO. In accordance with Eqs. (3) and (4), replacing the air with a solvent in an IO material should result in an increase in $n_{eff}$, which subsequently causes a shift in the PBG by structural interference, as described by the Bragg-Snell relation. For normal incidence, the Bragg-Snell relation in Eq. (2) can be reduced to:

$$\lambda_{min} = 2d_{111}n_{eff} \qquad (5)$$

This relationship between $\lambda_{min}$ and $n_{eff}$ is clearly observed in Figs. 2 (c) and (d), and the position of the transmission minimum red-shifts for solvents possessing larger refractive indices. In addition to air ($n_{sol}$ = 1), the solvents tested include: methanol ($n_{sol}$ = 1.329), ethanol ($n_{sol}$ = 1.361), dimethyl carbonate ($n_{sol}$ = 1.393), tetrahydrofuran ($n_{sol}$ = 1.407), dichloromethane ($n_{sol}$ = 1.424), ethylene glycol ($n_{sol}$ = 1.432) and toluene ($n_{sol}$ = 1.495)[75].

The simulated band structure diagrams for a $TiO_2$ IO (D = 400 nm) with $\varphi_{IO} = 0.18$ and $SnO_2$ IO (D = 384 nm) with $\varphi_{IO} = 0.26$ are shown in Figs. 2 (a) and (b), respectively. Band structure diagrams are shown only in the Γ – M direction to illustrate the pseudo-photonic bandgap created by adding a solvent to the IO network. Full band diagrams for $TiO_2$ and $SnO_2$ IOs in other solvents can be found in the Supporting Information Figs. S12 and S13. Figures 2 (a) and (b) display the band structure for IOs in air, methanol, and toluene with the p-PBG highlighted and shaded in black, red and purple segments, respectively. These shaded regions displaying the projected p-PBG are replicated in Figs. 2 (c) and (d) correlating theoretical and experimental predictions of the transmission minimum, because the fill fraction could be tuned to match that predicted by detailed experimental analysis. Inset in Figs. 2 (a) and (b) is a schematic representation of the IO structure used to simulate the band structure in each case. Figure 2 (b) shows the $\varphi_{IO} = 0.26$ case for the $SnO_2$ IO, whereby the IO material fully occupies (dense infilling) all interstitial sites. The schematic diagram inset in Fig. 2 (a) shows a less than optimal fill fraction ($\varphi_{IO} = 0.18$), where the IO material occupies *all* interstitial sites, but in a less dense, porous manner.

Combining Eq. (5) with Eq. (4), the Bragg-Snell relation can be given in terms of the Parallel model for $n_{eff}$:

$$\lambda_{min} = 2d_{111}(n_{IO}\varphi_{IO} + c(1 - \varphi_{IO})) \quad (6)$$

Likewise, a combination of Eq. (5) with Eq. (3) provides a Bragg-Snell relation in terms of the Drude model for $n_{eff}$:

$$\lambda^2_{min} = 4d^2_{111}(n^2_{IO}\varphi_{IO} + n^2_{sol}(1 - \varphi_{IO})) \quad (7)$$

Using Eq. (5), a plot of $\lambda_{min}$ vs $n_{eff}$ yields a linear plot through the origin with no intercept ($\gamma_1$). The slope ($\mu_1$) can be used to compute a value for $d_{111}$. This method requires a known value for $n_{eff}$, and assumes a perfectly filled IO structure with $\varphi_{IO} = 0.26$. Some reports[53-56] have suggested that inverse opals feature a less than an optimal filling fraction. A variation to the filling fraction $\varphi_{IO}$ of an IO would prevent an accurate determination of $n_{eff}$ via Eqs. (3) and (4). We applied the Parallel model to the spectral data in Figs. 2 (c) and (d) for $TiO_2$ and $SnO_2$ IOs, respectively. In the case of the $SnO_2$ IO, the data fits very well with the Parallel model, featuring a value for the intercept of $\gamma_1 = -7.14$ nm, very close to zero. The slope $\mu_1 = 574.49$ nm approximates the center-to-center distance of the $SnO_2$ IO pores, yielding a value of 351.72 nm using the standard interplanar spacing of $d = \sqrt{\frac{2}{3}} D$ which is in excellent agreement with $D$ = 354 nm calculated from angle-resolved spectra in air.

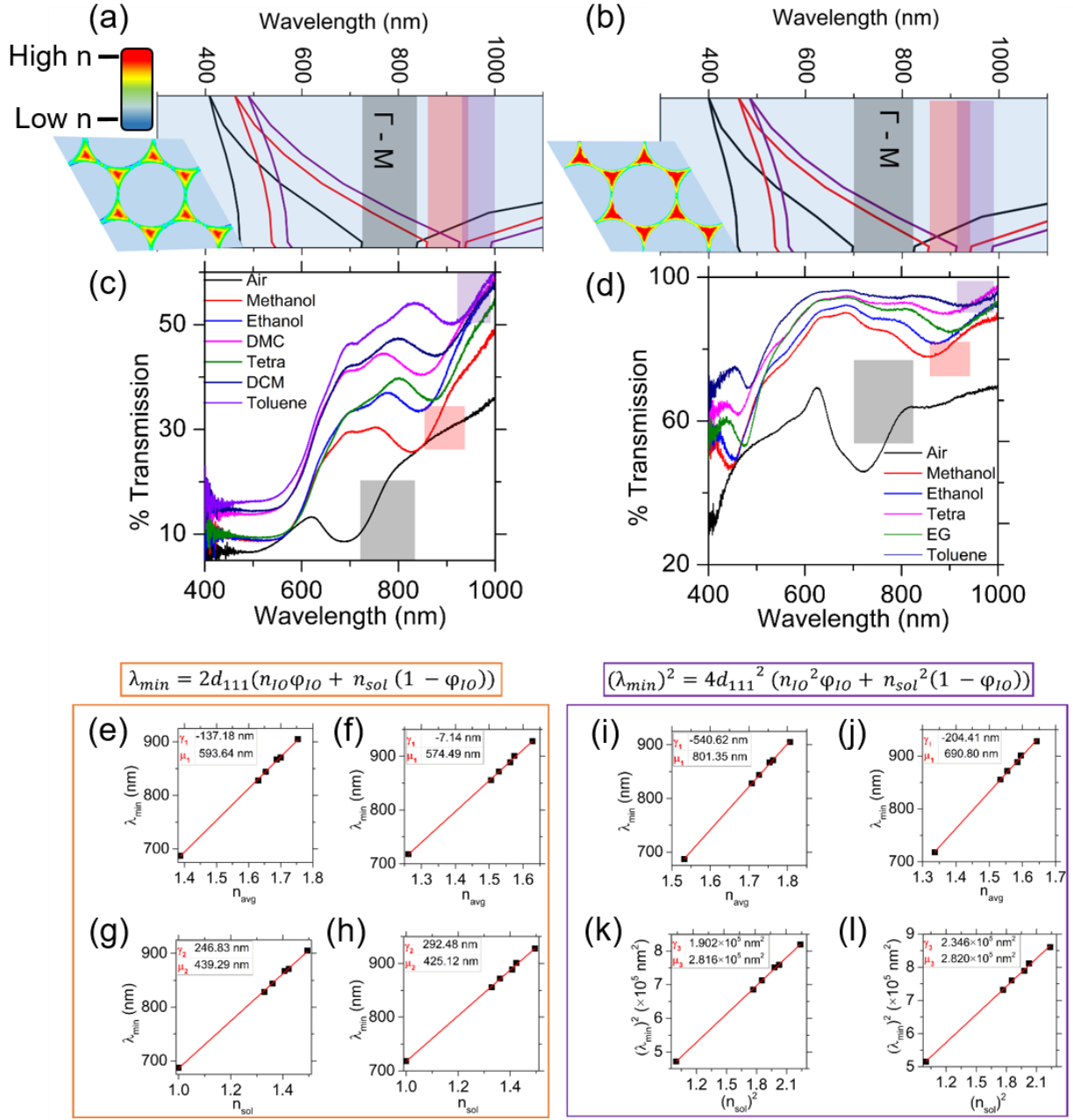

**Figure 2.** Band structure diagrams in the Γ – M direction showing projected photonic bandgap positions for (a) a TiO$_2$ IO ($\varphi_{IO}$ = 0.18) and (b) a SnO$_2$ IO ($\varphi_{IO}$ = 0.26) air (black), methanol (red) and toluene (purple). *Inset:* A schematic representation of the inverse opal structure used to model band structures, highlighting the differences in fill factor in each case. Higher index areas are shown in red (IO material) and lower index areas are shown in blue (air or solvent). Optical transmission spectra at normal incidence ($\theta$ = 0°) for (c) TiO$_2$ and (d) SnO$_2$ inverse opals infilled with a variety of solvents including methanol, ethanol, dimethyl carbonate (DMC), tetrahydrofuran (Tetra), dichloromethane (DCM), toluene and ethylene glycol (EG). Optical spectral analysis using the Parallel model for plots of $\lambda_{min}$ vs $n_{eff}$ for (e) TiO$_2$ IO and (f) SnO$_2$ IO and plots of $\lambda_{min}$ vs $n_{sol}$ for a (g) TiO$_2$ IO and (h) SnO$_2$ IO. Optical spectral analysis using the Drude model for plots of $\lambda_{min}$ vs $n_{eff}$ for (i) TiO$_2$ IO and (j) SnO$_2$ IO and plots of $(\lambda_{min})^2$ vs $(n_{sol})^2$ for a (k) TiO$_2$ IO and (l) SnO$_2$ IO.

Our data suggest that the Parallel model, with an assumed optimum $\varphi_{IO}$ = 0.26 provides an excellent estimation of $n_{eff}$ of the SnO$_2$ IO in a wide variety of solvents. In comparison, the data for the TiO$_2$ IO using the Parallel model do not conform as well as the SnO$_2$ IO. The intercept of $\gamma_1$ = −137.18 nm for the TiO$_2$ data is far from the ideal zero value. This significant deviation suggests that the Parallel model with an assumed $\varphi_{IO}$ = 0.26 is inaccurate for describing the behaviour of the TiO$_2$ IO. Any attempts to calculate a center-to-center pore distance from the slope would not provide an accurate estimation; the value of $\gamma_1$ would act to heavily skew $\mu_1$. Figures 2 (g) and (h) present the corresponding data analysis for the same TiO$_2$ and SnO$_2$ IO using the Drude model. In both cases it is immediately obvious from the very large negative values of the intercepts, −540.62 nm and −204.41 nm for TiO$_2$ and SnO$_2$ respectively, that this model with $\varphi_{IO}$ = 0.26 does not accurately predict $n_{eff}$ of either structure. The Parallel model provides a more accurate estimate of the effective refractive index of IO structures, particularly in the case of the SnO$_2$ IO.

To explain the differences between the experimental results and theoretical predictions, it is common practice to compute a custom value for the fill factor to better fit the data[53,57-61], but such of the reported values are so low that they are meaningless when compared to detailed electron microscopy analyses of the structures. A custom value for the fill factor can be obtained from Eqs. (6) and (7). Considering the Parallel model as shown in Eq. (6), a plot of $\lambda_{min}$ versus $n_{eff}$ should feature slope $\mu_2 = 2d_{111}(1 - \varphi_{IO})$ and an intercept $\gamma_2 = 2d_{111}n_{IO}\varphi_{IO}$. It follows that the centre-to-centre pore distance (D) and the volume fill factor of the inverse opal ($\varphi_{IO}$) can be found using:

$$D = \frac{\mu_2 n_{IO} + \gamma_2}{2d_{111}n_{IO}} \qquad (8)$$

$$\varphi_{IO} = \frac{\gamma_2}{\mu_2 n_{IO} + \gamma_2} \qquad (9)$$

Plots of $\lambda_{min}$ versus $n_{eff}$ for a typical TiO$_2$ and SnO$_2$ IO can be found in Figs. 2 (e) and (f), respectively. For the TiO$_2$ IO, using $\mu_2$ = 439.29 nm and $\gamma_2$ = 246.83 nm, a value for $D$ = 330 nm using the standard interplanar spacing ($d = \sqrt{\frac{2}{3}} D$) or 466 nm using the reduced interplanar spacing ($d = \sqrt{\frac{1}{3}} D$). Compared to the SEM measured mean of $D$ = 426 nm, the standard spacing underestimates the center-to-center to pore distance, whereas the reduced spacing provides a closer prediction to the measured value. The value for the fill factor of the TiO$_2$ IO was found as 18.4%, using Eq. (7). This represents a less than optimal filling (material with porosity) of the TiO$_2$ IO compared to the expected 26% for ideal inverse opal structures. For the SnO$_2$ IO, applying Eq. (8) to a slope of $\mu_2$ = 425.12 nm and $\gamma_2$ = 292.48 nm yields a value of $D$ = 350 nm for $d = \sqrt{\frac{2}{3}}\,D$ and $D$ = 494 nm for $d = \sqrt{\frac{1}{3}}\,D$. Comparing these values to the SEM measured mean of 384 nm, the standard spacing of $d = \sqrt{\frac{2}{3}}\,D$ provides a closer estimate to the center-to-center pore distance evaluated from microscopy images. This agrees with angle-resolved spectral measurements for the SnO$_2$ IOs in air, where the standard spacing more accurately predicts a value for $D$. The corresponding fill factor for the SnO$_2$ IO using Eq. (9) is 25.5% and this suggests that the IO is completely filled with dense SnO$_2$.

Using the Parallel model to calculate $n_{\text{eff}}$, TiO$_2$ IOs follow Bragg-Snell theory for photonic crystals using a reduced interplanar spacing ($d = \sqrt{\frac{1}{3}}\,D$), which provides a better estimate for the center-to-center pore distance. Conversely, SnO$_2$ IOs report a near-perfect volume filling fraction of IO material, with the standard interplanar spacing ($d = \sqrt{\frac{2}{3}}\,D$) yielding closer estimates to the measured center-to-center pore distance. This somewhat surprising finding suggests that the material in sub-optimally filled IO structures is correctly described using a reduced interplanar spacing when interacting with incident light. Previously, through analysis of the optical spectra of TiO$_2$ IOs with controlled variation in the angle of incidence, we suggested[52] a preference for light interaction with tetrahedral sites in the unit cell of IO structures, introducing the concept of a reduced interplanar spacing, which accounts for a net reduction in fill factor ($\varphi_{\text{IO}}$) where all voids are filled with less dense material overall. Other works have also suggested a reduced interplanar spacing[17,51]. It may be the case that perfectly filled IO structures (all voids filled with dense material) adopt the standard interplanar spacing, interacting with the full unit cell of tetrahedral and octahedral sites in IO structures.

Regarding the Drude model at normal incidence as described in Eq. (7), a plot of $(\lambda_{\min})^2$ versus $(n_{\text{sol}})^2$ will have a slope $\mu_3 = 4d_{111}^2(1 - \varphi_{\text{IO}})$ and an intercept $\gamma_3 = 4d_{111}^2\varphi_{\text{IO}} n_{\text{IO}}^2$. These plots can be seen in Figs. 2 (i) and (j) for a TiO$_2$ and SnO$_2$ IO, respectively. Therefore, the center-to-center pore distance ($D$) and the volume fill factor of the inverse opal ($\varphi_{\text{IO}}$) can also be computed for the Drude model using:

$$D = \sqrt{\frac{\mu_3\, n_{\text{IO}}^2 + \gamma_3}{4 d_{111}^2 n_{\text{IO}}^2}} \qquad (10)$$

$$\varphi_{\text{IO}} = \frac{\gamma_3}{\mu_3\, n_{\text{IO}}^2 + \gamma_3} \qquad (11)$$

Applying Eq. (11), a fill factor of just 9.8% for a $TiO_2$ IO and 17.8% for a $SnO_2$ IO is found via the Drude model. Using Eq. (10), the associated center-to-center pore distances assuming $d = \sqrt{\frac{2}{3}} D$, are $D$ = 520 nm for $TiO_2$ IOs and $D$ = 543 nm for $SnO_2$ IOs. Both of these estimates for the center-to-center distances are much larger than observed from microscopy inspection. In fact, these estimates are larger than the periodicity present in the initial opal template. Supposing instead $d = \sqrt{\frac{1}{3}} D$, we now have $D$ = 437 nm for the $TiO_2$ IO and $D$ = 456 nm for the $SnO_2$ IO. For the $TiO_2$ IO, 437 nm is reasonably close to the SEM measured mean of 426 nm. However, for the $SnO_2$ IO, 456 nm far exceeds the SEM measured mean of 384 nm, proving that the Parallel model is appropriate for fully filled IOs.

Comparing the results of these two models the Parallel model (Eq. (6)) provides a much better estimate for the effective refractive index compared to the Drude model (Eq. (7)) for IO-type photonic crystals in air, in general. Centre-to-centre pore distances estimated via the Parallel model match SEM measurements, particularly in the case of the $SnO_2$ IOs. In the case of the $SnO_2$ IO, the 25.5% filling fraction found using the Parallel model is almost exactly matched with the 26% expected from the inverse configuration of a stack of spheres coordinated in an FCC (111) geometry.

**Photonic bandgap characteristics of solvent-filled inverse opals**

To accurately determine the nature of the PBG based on index contrast and the degree of infilling, we extended our analysis light transmission at normal incidence when solvents were infiltrated, to angle-resolved measurements. In order to guarantee an accurate collection of data for such a system, measures were taken to ensure the correct angle of incidence was recorded for any given data set, summarized in the Supporting Information and shown in Fig. S4.

Shown in Figs. 3 (a-c) are the optical transmission spectra of photonic crystal structures in ethanol, showing a 350 nm polystyrene opal, a $TiO_2$ IO and a $SnO_2$ IO (both fabricated from a $D$ = 500 nm opal template), respectively. In each case, the optical spectra are recorded at normal incidence, 5°, 10°, 15°, and 20°. The transmission minimum associated with the PBG and the spectra profile of the IOs are symmetrically consistent for each angle of incidence either side of normal incidence, indicating a high-quality IO with uniform thickness and structure, see Supporting Information Fig. S4 – S9 for additional spectra.

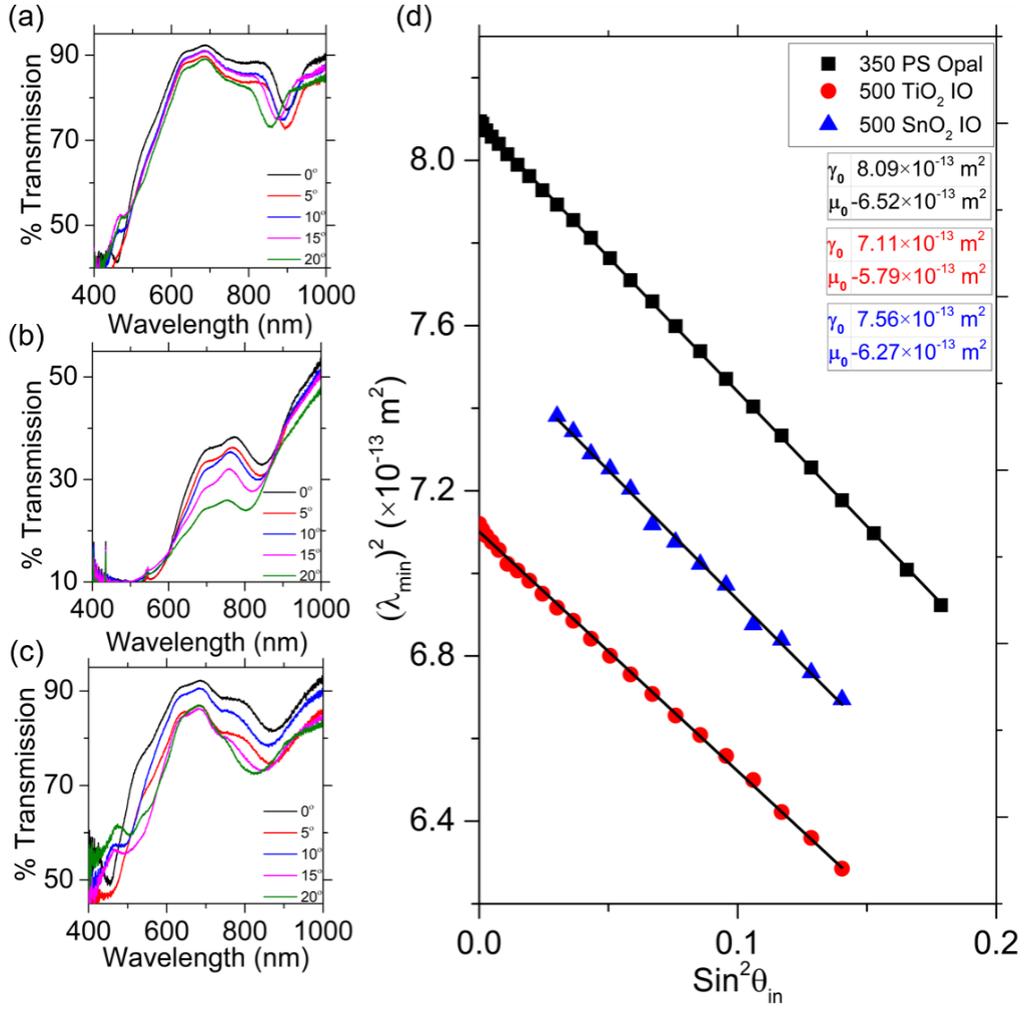

**Figure 3.** Optical transmission spectra recorded at 0°, 5°, 10°, 15° and 20° AOI for (a) a $D = 350$ nm diameter polystyrene opal, (b) $TiO_2$ and (c) $SnO_2$ IOs (fabricated from $D = 500$ nm polystyrene opal templates) immersed in ethanol. (d) Corresponding Bragg-Snell analyses of $\lambda^2$ versus $\sin^2\theta$ for the photonic crystals in ethanol.

We would expect the Bragg-Snell relation in Eq. 2, applied to the inverse opal in air, to also describe the behavior of the optical transmission spectrum in a given solvent. Therefore, for a plot of $(\lambda_{min})^2$ versus $\sin^2\theta$ with an expected slope $\mu_0 = 4d_{111}^2 n_{sol}^2$ and an intercept $\gamma_0 = 4d_{111}^2 n_{eff}^2$, one would expect to find:

$$\frac{(n_{sol}^2)\lambda_0}{\mu_0} = n_{eff}^2 \qquad (12)$$

Bragg-Snell data of $(\lambda_{min})^2$ versus $\sin^2\theta$ can be seen for the $D = 350$ nm PS opal, $D = 500$ nm $TiO_2$ and $SnO_2$ IOs in ethanol through Fig. 3 (d). A clear linear regression is observed in each case, as expected from Eq. 2. Next, we obtained full transmission data for both IOs in all solvents, and a collection of plots of $(\lambda_{min})^2$ versus $\sin^2\theta$ are shown in Figs. 4 (a) and (b). We find that the linear relation described by Eq. 2, is still retained even when the IO network in filled with a solvent. A larger range of solvents was tested for the $TiO_2$ IO as solvents with a higher refractive index maintained a greater index contrast, and $TiO_2$ IOs ($n = 2.49$), supporting a stronger inhibition of light propagation (lower transmittance) around the PBG compared to $SnO_2$ in the same solvents. This PBG-quenching effect for the SnO$_2$ IO ($n$ = 2.01) can be seen in Fig. 2 (d), with a noticeable decrease in depth of the transmission minimum in the case of toluene ($n$ = 1.495). Introducing higher refractive index solvents reduces the refractive index contrast between the periodic crystalline material and the medium filling the pores in the IO structure. Nevertheless, transmission data with a varying angle of incidence could be recorded accurately for multiple solvents for both IO materials.

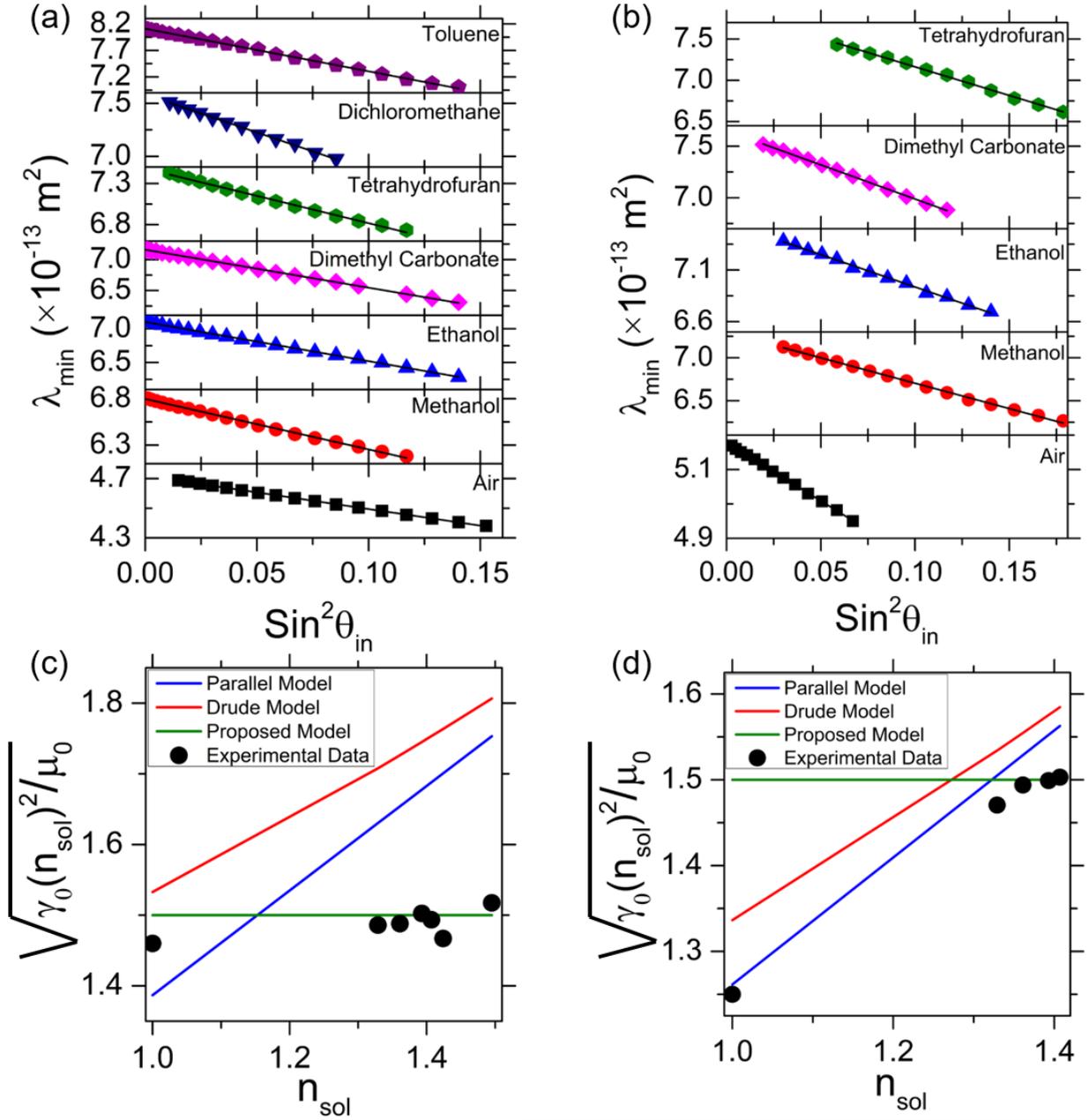

**Figure 4.** Bragg-Snell plots of ($\lambda_{min}$)$^2$ vs sin$^2\theta$ for (a) TiO$_2$ and (b) SnO$_2$ IOs shown for air, methanol, ethanol, dimethyl carbonate, tetrahydrofuran, dichloromethane, and toluene-filled pores. Analysis of the Bragg-Snell data for (c) TiO$_2$ and (d) SnO$_2$ IOs, showing the results of the data (in black circles) as per Eqs. (12) and (14), the refractive index of the substrate $n_{\mathrm{sub}}$ is shown as the green line and the effective refractive indices calculated via the Drude (red) and the Parallel (blue) models are also shown for an optimally filled IO structure ($\varphi_{IO}$ = 0.26).

Our measurements on solvent-filled IOs reveal that the standard analytical treatment via Bragg-Snell relations *must account for the substrate.* Figures 4 (c) and (d) contain the results of the Bragg-Snell analysis of the optical transmission spectra described in Eq. (12) for $TiO_2$ and $SnO_2$ IOs, respectively. We find an obvious disagreement between the expected results from Eq. (12) and the observed results from careful experimental analysis. The data show that experimentally calculated values for the $n_{eff}$ obtained from angle-resolved transmission spectra, remain invariant with the corresponding increase in solvent refractive index. In fact, the value (calculated via the left-hand side of Eq. (12)) is relatively constant at ~1.5 for both IO structures in the presence of a solvent, in spite of the different refractive indices of $TiO_2$ and $SnO_2$. In Figs. 2 (c) and (d), we found a clear shift in the position of the PBG when the IO was filled with solvent. This magnitude of the red-shift increases with higher values for the solvent refractive index $n_{sol}$. As such, a greater $n_{eff}$ of the system can be measured spectrally, even at normal incidence.

To explain the differences between the expected response of a solvent-filled IO photonic crystal and those measured using angle-resolved transmission, we propose a model to describe the light interaction with an IO on a substrate when infilled with a solvent. Noting the near-constant value of 1.5 obtained experimentally and its similarity to the refractive index of the glass substrate (*n* = 1.5), we suggest that treatment of the Bragg-Snell data typically used for interpretation of IO spectral response, per Eq. (12), actually acts to calculate the refractive index of the substrate ($n_{sub}$), specifically when the IO is immersed in a solvent and the $n_{eff}$ of the IO layer exceeds that of the substrate layer. This scenario would likely not occur for most IO materials in air where typically the effective refractive index of the IO layer would not exceed the refractive index of the substrate layer. In terms of conceptualizing the difference to the light interaction, Figs 5 (a) and (b) present the standard approach and the proposed treatment of the data in solvents, respectively. In the derivation of the Bragg-Snell relation, the internal angle $\varphi$ is used and the presence of the substrate layer is normally ignored as the conditions for constructive interference are satisfied in the IO layer, just as shown in Fig. 5 (a). One method of rationalizing the experimental results observed in the solvent case would be to adopt the internal angle $\varphi'$, as seen in Fig. 5 (b). Incorporating this change into Eq. (2) yields the following relation:

$$\lambda^2 = 4d_{111}^2 n_{eff}^2 \left(1 - \frac{n_{sol}^2}{n_{sub}^2} \sin^2 \theta \right) \qquad (13)$$

Looking at Eq. (13), at normal incidence, the optical properties of the PBG behave normally. However, the spectral response as a function of the angle of incidence is different. Specifically, using this relation, a plot of $\lambda^2$ vs $\sin^2\theta$ would now feature a slope $\mu_4 = -4d_{111}^2 n_{avg}^2 \frac{n_{sol}^2}{n_{sub}^2}$ and an intercept $\gamma_4 = 4d_{111}^2 n_{avg}^2$. Therefore, Eq. (12) now is of the form:

$$\frac{(n_{sol}^2)\gamma_4}{\mu_4} = n_{sub}^2 \qquad (14)$$

This gives an excellent fit with the experimentally determined values for both IO structures in solvents, with the approximate value for the refractive index of the glass substrate ($n$ = 1.5) coinciding with near-constant value of 1.5 found experimentally using Eq. (12).

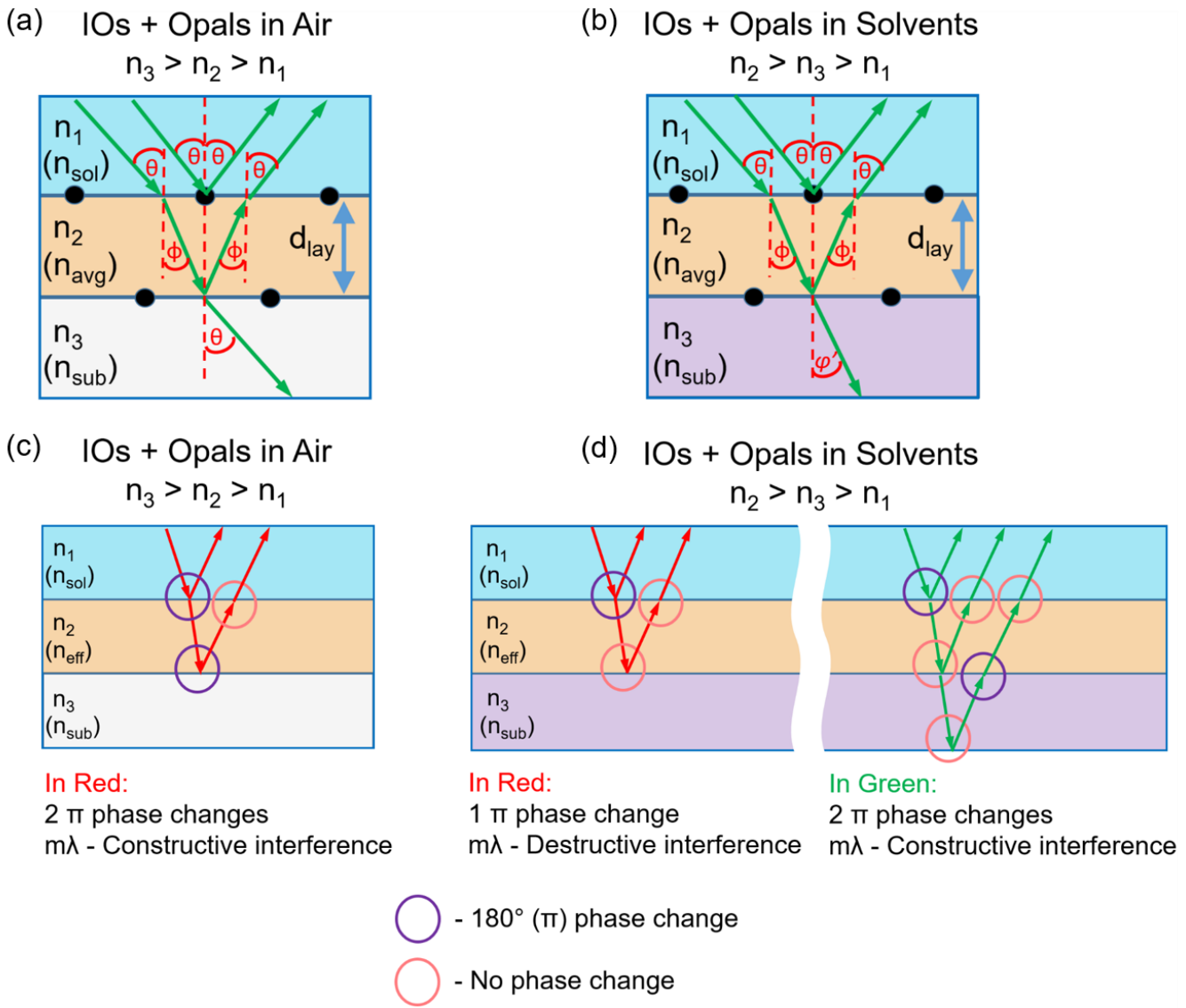

**Figure 5.** Schematic representations of the (a) standard treatment of the internal angle for IO structures for no substrate interaction and (b) proposed treatment of the internal angle for IO structures in solvents considering a substrate interaction. Schematic diagram showing the proposed difference between the constructive interference conditions for (a) an inverse opal in air with $n_{sub} > n_{eff} > n_{sol}$ and (b) an inverse opal in a solvent with $n_{eff} > n_{sub} > n_{sol}$. The constructive interference condition corresponding to an integer number of wavelengths (matching the Bragg-Snell condition for opals and inverse opals) is satisfied for two separate π phase changes.

We posit that the explanation for this change to the Bragg-Snell relation can be found by considering the changes to the constructive interference condition of the light upon the addition of the solvent. We note that all earlier findings in this paper that address the true response of IOs in solvents remain valid. Normally, the Bragg-Snell relation is derived for a single unit cell of material in an IO layer, presumably since the constructive interference effect for the entire system is satisfied for $n_{\text{eff}} < n_{\text{sub}}$. This can be seen in Fig. 5 (c), where two π phase changes to the light are experienced upon reflection from the top surface of the substrate layer. For $n_3 > n_2 > n_1$, the condition for constructive interference is as follows:

$$m\lambda = 2n_2 d_{\text{lay}} \cos\theta_2 \qquad (15)$$

In this scenario, the distance $d_{\text{lay}}$ corresponds to the entire thickness of the second layer of refractive index $n_2$. This condition is valid for integer values of *m*. This integer condition for the entire system matches the constructive interference condition assumed when deriving the Bragg-Snell relation using the unit cell dimension of the IO. For an IO material of refractive index $n_{\text{eff}}$, with an internal angle $\theta_2$, constructive interference is given by:

$$m\lambda_{111} = 2d_{111} n_{\text{eff}} \cos\theta_2 \qquad (16)$$

In both cases for constructive interference described by Eqs. (15) and (16), the light will constructively interfere for some integer value (*m* = 0, 1, 2, …etc.) multiples of the wavelength.

In the solvent-filled IO structure, as depicted in Fig. 5 (d), there is no longer a phase change upon reflection from the top surface of the substrate later. For $n_2 > n_3 > n_1$, the condition for constructive interference now becomes:

$$(m + \tfrac{1}{2})\lambda = 2n_2 d_{\text{lay}} \cos\theta_2 \qquad (17)$$

Therefore, considering the three-layer system described in Fig. 5, the condition for constructive interference should be satisfied as a product of some half-integer value of the wavelength. However, analysis of the data collected at normal incidence in Figs. 2 (a) and (b) shows that the condition detailed in Eq. (16) describes the behavior of the PBG at normal incidence. Thus, integer values better model the system.

If a reflection from the bottom of the substrate layer is now considered, as shown in green in Fig. 5 (d), a second π phase change is now incorporated into the interaction. This would return the condition for constructive interference for the three-layer system to that described in Eq. (15). Essentially, this would imply that the *mλ* condition required for reflection by the individual unit cell layers in the IO material must be matched by corresponding *mλ* condition for the entire three-layer system. This supposition implies that the internal angle required for the Bragg-Snell model may be that of the substrate layer, which would fit with the proposed change to the Bragg-Snell relation for IOs in solvents suggested in Eq. (13).

## Conclusions

Monitoring the angle-dependent characteristics to light transmission through $TiO_2$ and $SnO_2$ inverse opals has uncovered several unique features that, taken together, shed light on the nature of the photonic bandgap when the index contrast is tuned by infiltration in a range of common solvents. The behavior of the photonic bandgap for inverse opal structures in solvents appears to display a clear deviation from the expected optical response, yet nonetheless, it conforms to a predictable result following subtle changes to the governing Bragg-Snell relation. When applying the standard Bragg-Snell approach commonly used for analyzing opal photonic crystals, we consistently find a discrepancy that depends on the index contrast between the solvent and IO material, the index of the transparent substrate, and the degree of infilling. This approach solves the issue related to fill fraction, commonly reported to inconsistently low in many investigations involving IO strucutres.

Estimations of the volume fraction of material present via solvent-induced PBG shifts at normal incidence, confirm a slight variance in the material filling for IO structures. $SnO_2$ inverse opals report a volume fraction of 25.5% close to the theoretical maximum of 26%. Slight infill errors appear to persist in $TiO_2$ inverse opals, with a volume fraction of 18.4% falling below the expected 26%. The $TiO_2$ inverse opal would appear to require a reduced interplanar spacing ($d = \sqrt{\frac{1}{3}} D$) to align with expectations from the theory. We suggest the possibility that the need for a reduced interplanar spacing arises from the suboptimal volume filling of the inverse opal structure. This reduced filling is quite specific and describes a less dense material filling all the voids as opposed to dense material materials filling some of the voids, which is corroborated by experimental data and photonic band structure modeling. Fully filled photonic crystal structures, as in the case of the $SnO_2$ inverse opal, appear to adhere to the expected interplanar spacing ($d = \sqrt{\frac{2}{3}} D$) for an FCC (111) structure, and in spite of difference in index contrast between both materials in several solvents, the degree of infilling correlates with the interplanar spacing of the IO structure. As such, reduced filling fractions modify the PBG directionality and bandwidth so that a modified unit cell is necessary to correctly fit that angle-dependent spectral response. Differences in the perceived interplanar spacing between the two materials would account for the contradictory appearance of the spectra at normal incidence, explaining why the photonic bandgap for $TiO_2$ inverse opals is located at shorter wavelengths than $SnO_2$ inverse opals.

To reconcile this observed difference between the theoretical expectations and experimental data, we put forward a conflict in constructive interference conditions for the system when an IO is filled with a solvent. Adding a solvent to the inverse opal material causes the effective refractive index of the inverse opal layer to exceed that of the substrate layer for most common materials and solvents. This has the effect of changing the constructive interference condition for the entire system from the integer-valued $m\lambda$ to the half integer-valued $(m + \frac{1}{2})\lambda$. Further interaction with the glass substrate layer would revert the constructive interference condition back to $m\lambda$. Modeling the Bragg-Snell relation using the internal angle of the glass substrate would effectively explain the near-constant value of 1.5 estimated for the refractive index by the experimental data.

This analysis comes with the unfortunate caveat that the center-to-center pore distance and effective refractive index of the inverse opal materials can no longer be independently calculated spectrally and detailed microscopy is always recommended, but the interpretation provides the first fully consistent correlation between fully and partially filled IOs, with tunable index contract in common solvents, and the nature of their PBG. Whatever the underlying reasoning for the disagreement between the theoretical and experimental data, it would seem likely from analysis of the optical spectra that the Bragg-Snell relation applied to an inverse opal in air does not directly model the behavior of the inverse opal system when immersed in a solvent, on a transparent substrate. This data provides the basis for understanding how structural color, defined by a PBG, of IOs on a substrate immersed in solvents can be extracted. Such information is critical to any application of an order porous IO material analyzed by spectroscopy in a liquid, which spans disciplines from electrochemical energy conversion and storage and soft photonics, to biological sensors to photocatalysts and solar cells.

**Acknowledgments**

We acknowledge support from the Irish Research Council Government of Ireland Postgraduate Scholarship under award no. GOIPG/2016/946. This work was supported by Science Foundation Ireland under contract no. 13/TIDA/E2761, 15/TIDA/2893 and 17/TIDA/4996. This publication has also emanated from research supported in part by a research grant from SFI under Grant Number 14/IA/2581. We also acknowledge funding from the Irish Research Council Advanced Laureate Award under grant no. IRCLA/2019/118.

**Author information**

Corresponding author: Colm O'Dwyer, email: c.odwyer@ucc.ie; Tel: +353 21 490 2732

Alex Lonergan[1], Changyu Hu[1], and Colm O'Dwyer[1,2,3,4]*

[1]School of Chemistry, University College Cork, Cork, T12 YN60, Ireland
[2]Micro-Nano Systems Centre, Tyndall National Institute, Lee Maltings, Cork, T12 R5CP, Ireland
[3]AMBER@CRANN, Trinity College Dublin, Dublin 2, Ireland
[4]Environmental Research Institute, University College Cork, Lee Road, Cork T23 XE10, Ireland

**Additional SEM images for TiO$_2$ inverse opals:**

Further SEM images for a typcial TiO$_2$ inverse opal can be seen in Fig. S1. Figures S1 (a) – (c) show the FCC (111) plane expected for an ordered inverse opal structure under various levels of magnification. Looking at the SEM images, it is clear that there are a small number of defects, particularly on the top layer surface. However, the long range order of the repeating strucuture is also maintained over large distances. At higher magnifications, subsequent layers can be observed through the pores in the top layer. These factors are indicative of a uniform inverse opal strcuture with a high degree of order. Figure S1 (d) is representative of the manner in which centre-to-centre pore distances were evaluated for inverse opal structures. As seen in Fig. S1 (d), there is a small degree of variation in centre-to-centre pore distances measured. Measurements were taken at various points and directions across the IO surface in order to accommodate for any anisotropy present in the IO structure. A size distribution of centre-to-centre pore distances was subsequently created using several hundred such measurements, as presented for a TiO$_2$ IO in Fig. 1 (g). Using this distribution, a mean centre-to-centre pore distance could be approximated for comparison with spectral data and a standard deviation value could be found to estimate the variance present in the centre-to-centre pore distances.

Exposure of the inverse opal structure to certain solvents had the effect of introducing limited surface deformation, as seen in Fig. S1 (e). The TiO$_2$ inverse opal shown in Fig. S1 (e) had been rigorously tested in methanol, ethanol, dichloromethane and toluene. The pristine nature of the inverse opal seems to have been lost, with visible damage throghout the strucuture. However in spite of the damage to the walls, the overall structure of the material is retained, with SEM images showing identical multilayer islands or domains of inverse opal material. Interestingly, the damaged structure still exhibits a sensible optical response, with little change to the optical results found. The SnO$_2$ IO did not display a similar structural degradation following testing with methanol, ethanol and dimethyl carbonate. This might identify toluene and dichloromethane as more damaging to the IO structure.

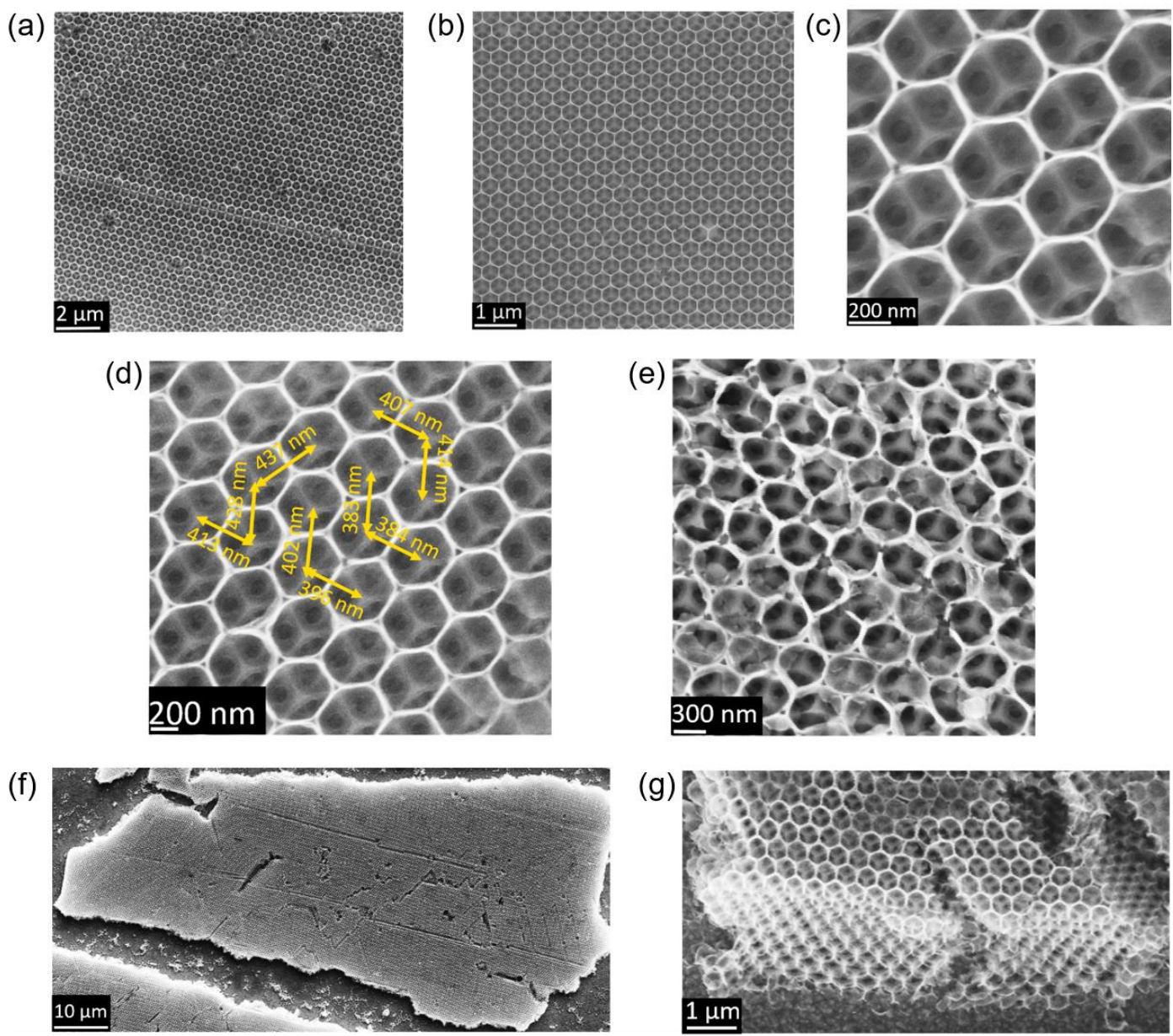

**Figure S1.** (a) – (c) SEM images of increasing magnification for a typical $TiO_2$ inverse opal, showing the expected geometry for an inverse opal. (d) Measured centre-to-centre pore distances in various directions for a $TiO_2$ inverse opal. (e) SEM image of a $TiO_2$ inverse opal following exposure to methanol, ethanol, dichloromethane and toluene. (f) Low magnification SEM image showing a typical domain of IO material and associated cracks. (g) SEM image showing the edge of an IO domain with a number of layers of IO material visible.

Figure S1 (f) features a low magnification SEM image, showing the size of a typical IO domain relative to the cracks in the material. For the domain surface featured, dimensions are approximately 90 × 40 µm. Several defects can be seen on the surface of the structure; the majority of the surface is composed of well-ordered IO material. The number of layers comprising the stack of IO material can be seen in Fig. S1 (g). The edge of an IO domain can be seen in the SEM image, with approximately 10 layers of IO material visible here. The order of the IO template can also be seen to continue throughout the structure from this image, confirming the expected three dimensional architecture for the inverse opal.

**Additional SEM images for SnO$_2$ inverse opals:**

SEM images of an SnO$_2$ inverse opal with increasing magnification can be seen in Fig. S2 (a) – (c). Much like in the case of the TiO$_2$ inverse opal, a high degree of long-range uniformity is observed for the inverse opal. Defects are present for the SnO$_2$ IO, however the majority of the surface is covered with the expected hexagonal geometry of the FCC (111) plane. In comparison with the TiO$_2$ IO, the SnO$_2$ IO appears to feature a higher number of thicker deposits on the top layer surface. Patches of thicker deposits appear more frequently as blotches on the surface of the SnO$_2$ IO. In spite of some minor differences, both IOs are structurally very similar, with a repeating geometry extending over large sections of the substrate surface. A low magnification SEM image of the SnO$_2$ IO can be seen in Fig. S2 (d). From this image, the extent of the coverage of the IO material across the substrate surface can be assessed. Cracks can be consistently observed between large domains of IO material, most likely casued by a contraction of the polymer material upon annealing at 450 °C[1,2]. Most importantly, the IO material can be seen to cover the majority of the surface of the substrate. Considering the diameter of the light beam (approx. 1 mm) interacting with the inverse opal, it is essential that the substrate features a high coverage of IO material. The SEM image of the SnO$_2$ IO shown in Fig. S2 (d), is an approximate representation of the area exposed to incident light.

Figure S2 (e) depicts the measurement of centre-to-centre pore distances for a typical SnO$_2$ IO. In comparison with the TiO$_2$ measurements, it would appear that centre-to-centre pore distances are slightly smaller in the case of the SnO$_2$ IO. This is reflected in the size distribution shown in Fig. 1 (i), whereby the mean centre-to-centre pore distance is measured as 384 nm for the SnO$_2$ IO, compared to 426 nm for the TiO$_2$ IO. This difference in pore diameter was consistently observed for SnO$_2$ IOs prepared via identical conditions to TiO$_2$ IOs. A typical domain of IO material can be seen in Fig. S2 (f) with approximate dimensions of 90 × 30 µm, similar in scale to TiO$_2$ IO domain sizes. Figure S2 (g) presents a tilted (45°) SEM image of an IO domain for a typical SnO$_2$ IO. For this particular domain, seven layers are visbile and appear to have sizes and geomtries consistent with the surface pores.

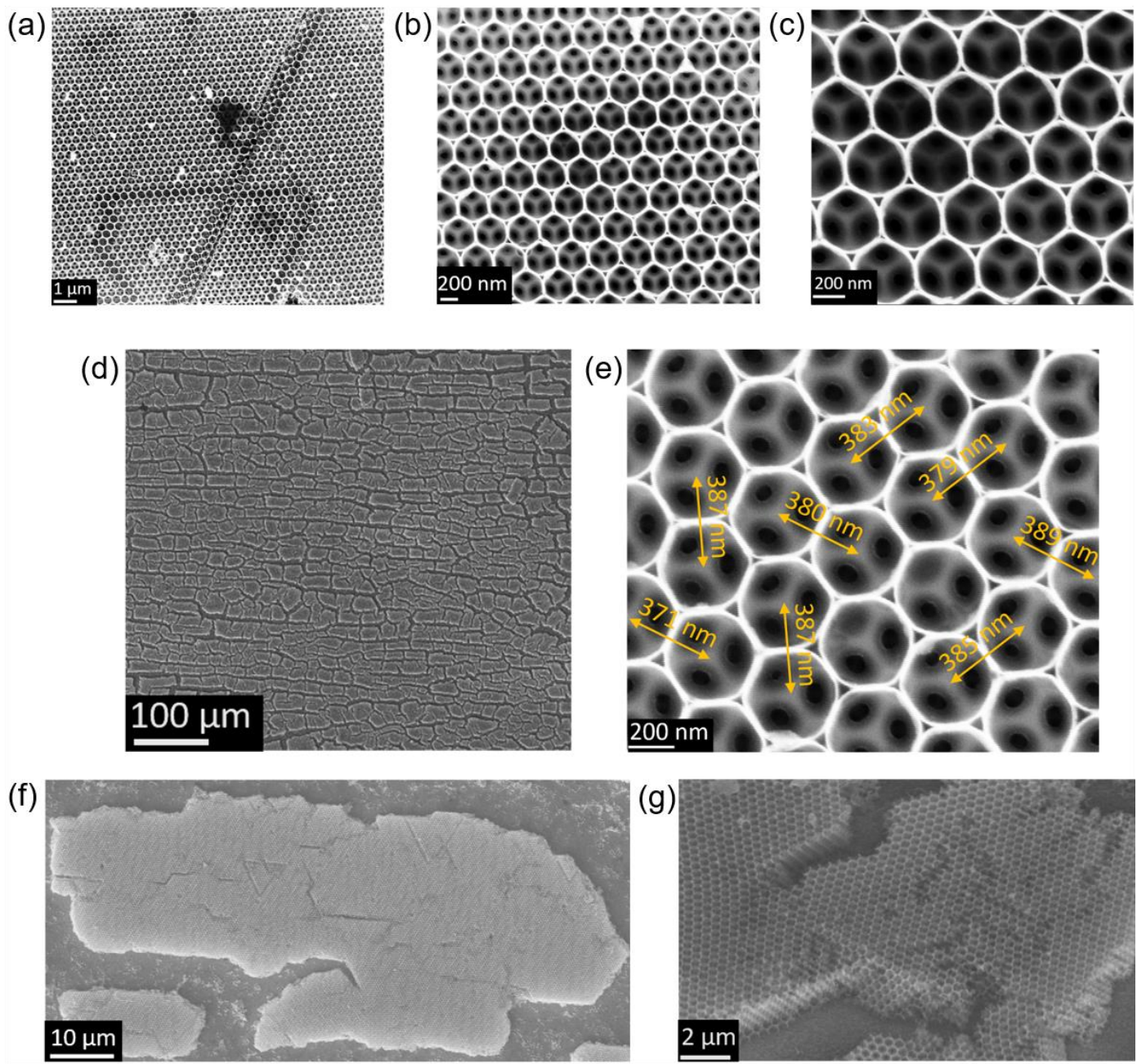

**Figure S2.** (a) – (c) SEM images of increasing magnification for a typical SnO$_2$ inverse opal, showing the expected geometry for an inverse opal. (d) Low magnification SEM image showing coverage of substrate surface with IO material. (e) Measured centre-to-centre pore distances in various directions for a SnO$_2$ inverse opal. (f) Low magnification SEM image showing a typical domain of IO material and associated cracks. (g) Tilted SEM image at 45 ° showing the edge of an IO domain with the number of layers of IO material visible.

**Raman scattering characterisation for inverse opals:**

The crystallinity and phase of the prepared IO materials is an important consideration for assessing the properties of the IO film. For optical analysis in particular, the refractive index of the IO material is crucial for estimation of the effective refractive index of the IO film. Using Raman analysis to accurately identify the material, allows the refractive index to be estimated via comparison with the literature. Figure S3 (a) displays the Raman spectrum for a typical TiO$_2$ IO, showing peaks at 150, 399, 524 and 645 cm$^{-1}$. These peaks show an excellent agreement with previous reports of anatase phase $TiO_2$[3,4] comparing favourably with reported $E_g$, $B_{1g}$, $A_{1g}$ and $E_g$ phonon modes of anatase phase $TiO_2$[5]. The Raman spectrum for a typical $SnO_2$ IO can be seen in Fig. S3 (b), alongside a normalised intensity comparison in Fig. S3 (c) showing the Raman peaks for the $SnO_2$ IO versus the peaks for the fluorine doped tin-oxide (FTO)-coated glass substrate. Bulk single crystal or polycrystalline $SnO_2$ has well documented Raman peaks of 478, 633 and 776 $cm^{-1}$ corresponding to the $E_g$, $A_{1g}$ and $B_{2g}$ Raman modes, respectively[6]. For the $SnO_2$ IO, Raman peaks can be seen at 353, 573, 637, 799 and 1107 $cm^{-1}$, whereas the FTO substrate shows peaks at 353, 475, 573, 799 and 1107 $cm^{-1}$. Both the $SnO_2$ IO and the FTO substrate display several characteristic peaks suggesting the presence of $SnO_2$. The peaks at 353 and 573 $cm^{-1}$ are typically attributed to the crystalline surface area associated with small grain sizes in nanostructured $SnO_2$[7,8].

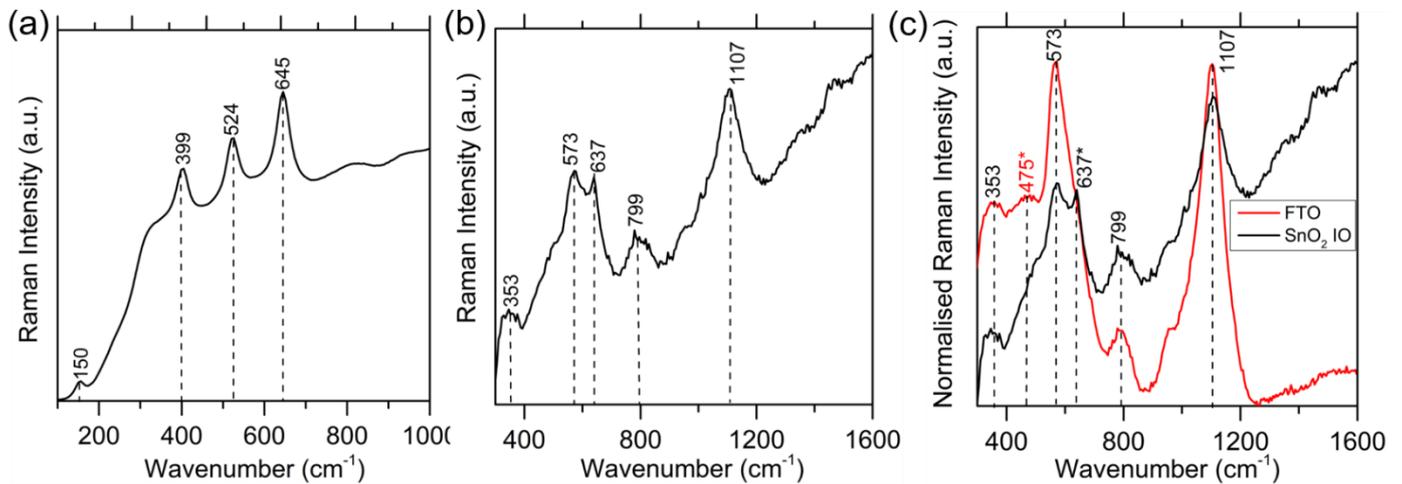

**Figure S3.** Raman scattering spectrum of a typical (a) $TiO_2$ and (b) $SnO_2$ IO prepared on FTO-coated glass. (c) A comparison of normalised Raman scattering spectra for a $SnO_2$ IO versus a bare FTO-coated glass substrate.

**Further optical analysis for inverse opals in various solvents:**

Samples were held vertically in a glass cylinder containing the solvent via a 3D printed removable grip. Data was collected symmetrically in both directions about the horizontal axis of the IO sample. This can be seen illustrated in Figs. S4 (a), (b) and (c). Figure S3 (a) displays the system with light at normal incidence. Figures S4 (b) and (c) show the direction and position of the sample with respect to the clockwise and anticlockwise direction about the central axis.

**Ethanol**

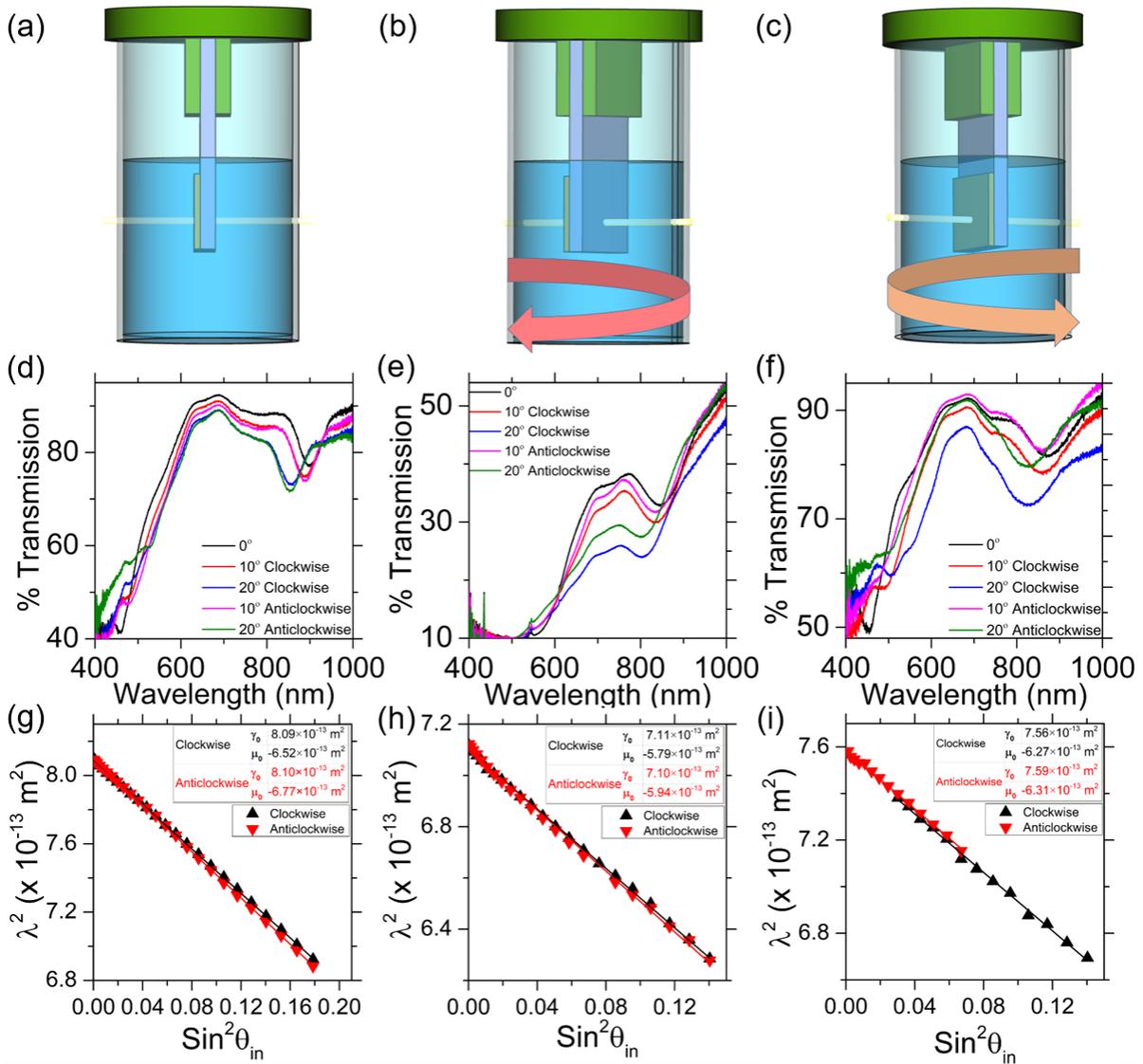

**Figure S4.** Schematic diagram showing the experimental set-up for testing inverse opals in solvents at (a) normal incidence, at an angle rotated in the (b) (forward) clockwise and (c) (backward) anticlockwise direction. Optical transmission spectra for (d) a $D = 350$ nm diameter polystyrene opal, (e) $TiO_2$ and (f) $SnO_2$ IOs (fabricated from $D = 500$ nm polystyrene opal templates), showing spectral symmetry in the clockwise and anticlockwise direction. Corresponding Bragg-Snell analyses of $\lambda^2$ versus $Sin^2\theta$ for the clockwise and anticlockwise directions showing slopes and intercepts as per Eq. (2) for ethanol infilled into (g) a 350 nm diameter polystyrene opal, (h) $TiO_2$ and (i) $SnO_2$ IOs.

The high spectral symmetry present in the transmission spectra for $D = 350$ nm polystyrene opals, $D = 500$ nm $TiO_2$ IOs and $D = 500$ nm $SnO_2$ IOs immersed in ethanol is displayed in Figs. S4 (d) – (f), respectively. The positions of the transmission minima are matched for both the forward (clockwise) and backward (anticlockwise) directions about normal incidence. The corresponding Bragg-Snell analyses further support the high symmetry observed in data collection. For each plot of $\lambda^2$ versus $Sin^2\theta$, the intercepts and slopes for both directions about the central axis, as per Eq. 2 in the main text, are very closely matched. The high level of data symmetry present would help to eliminate any possible errors associated with experimental set-up, ensuring that the (111) close packed plane is coincident with the 0° position used by the angle-resolved set-up. This is particularly important for conclusions drawn from the solvent-angle data in the main text, see Figs. 4 (c) and (d).

The optical response of inverse opal materials to variation of the angle of incidence upon immersion in a solvent is displayed for a larger number of solvents through Figs. S5 – S9. The data presented here is analogous to the data presented for the ethanol case in Fig. S4. Figures S5 – S9 correspond to accompanying datasets for methanol, dimethyl carbonate, tetrahydrofuran, dichloromethane and toluene, respectively. The results of the analysis of this solvent data, as per Eqs. (12) and (14), can be seen summarised and included in the main text in Figs. 4 (c) and (d).

As a general comment on the appearance of the spectra, a clear transmission minimum is present for each spectrum, shifting to lower wavelengths with increasing angle of incidence. The position of the transmission minimum at normal incidence is influenced by the refractive index of the solvent and thus, the effective refractive index of the IO film. This behaviour is identical to the ethanol spectra detailed in Fig. S4. In each instance, the position of the transmission minimum is symmetric in the clockwise and anticlockwise positions about normal incidence. This would indicate accurate data measurement for every solvent case. The Bragg-Snell data follows this trend, with values for the slopes and intercepts of the graphs ($\lambda_{min}$ vs $\text{Sin}^2 \theta_{in}$) being almost identical in the clockwise and anticlockwise directions. All of this would indicate little error (< 1°) in the assumed angle between incident light and the normal to sample surface.

**Methanol**

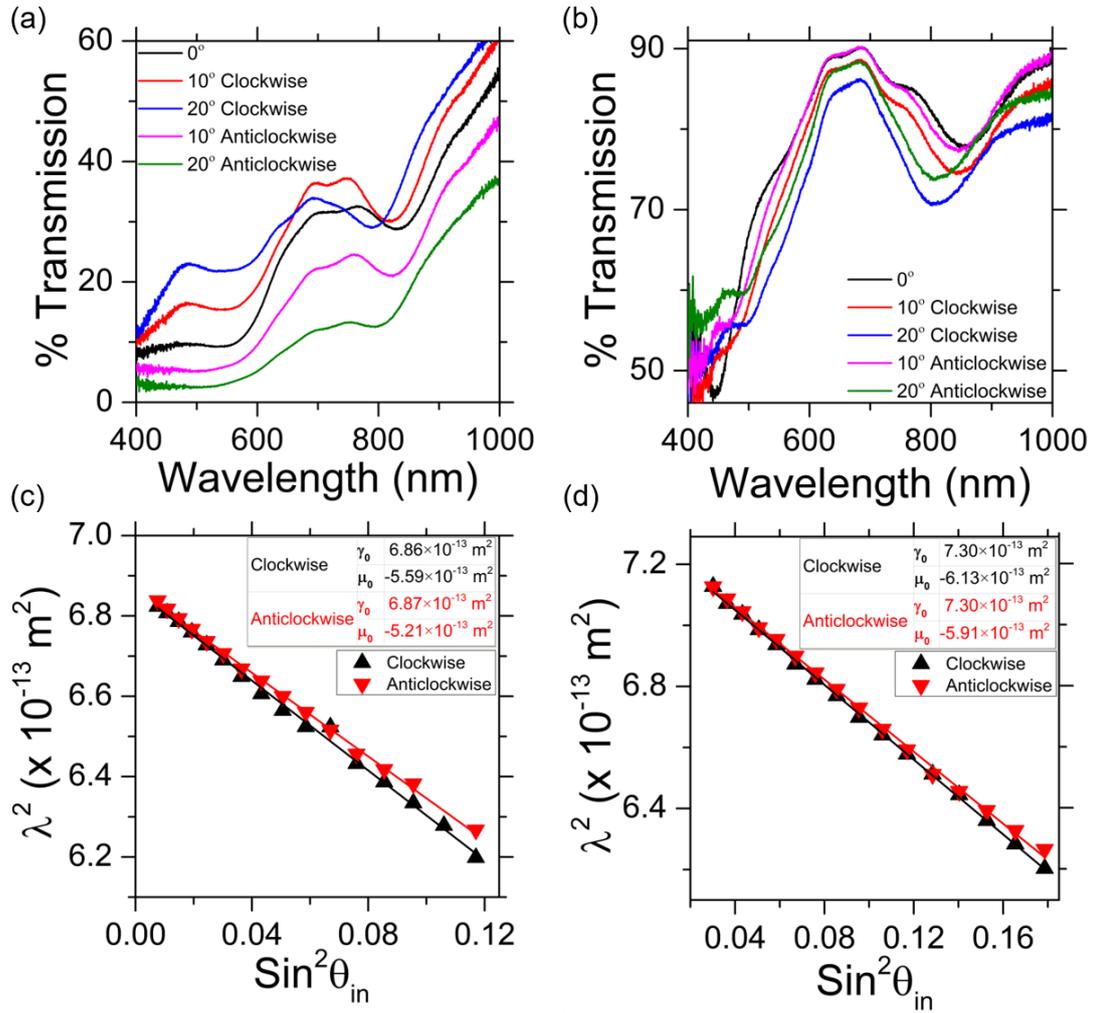

**Figure S5.** Optical transmission spectra for inverse opals immersed in methanol at normal incidence, 10° and 20° (both clockwise and anticlockwise) for a (a) $TiO_2$ and (b) $SnO_2$ IO. Corresponding Bragg-Snell plots of $\lambda_{min}^2$ vs $Sin^2 \theta_{in}$ in intervals of 1° for both a clockwise and anticlockwise direction about the vertical axis, showing a slope and an intercept value as defined via Eq. (2), for a (c) $TiO_2$ and (d) $SnO_2$ IO.

**Dimethyl Carbonate**

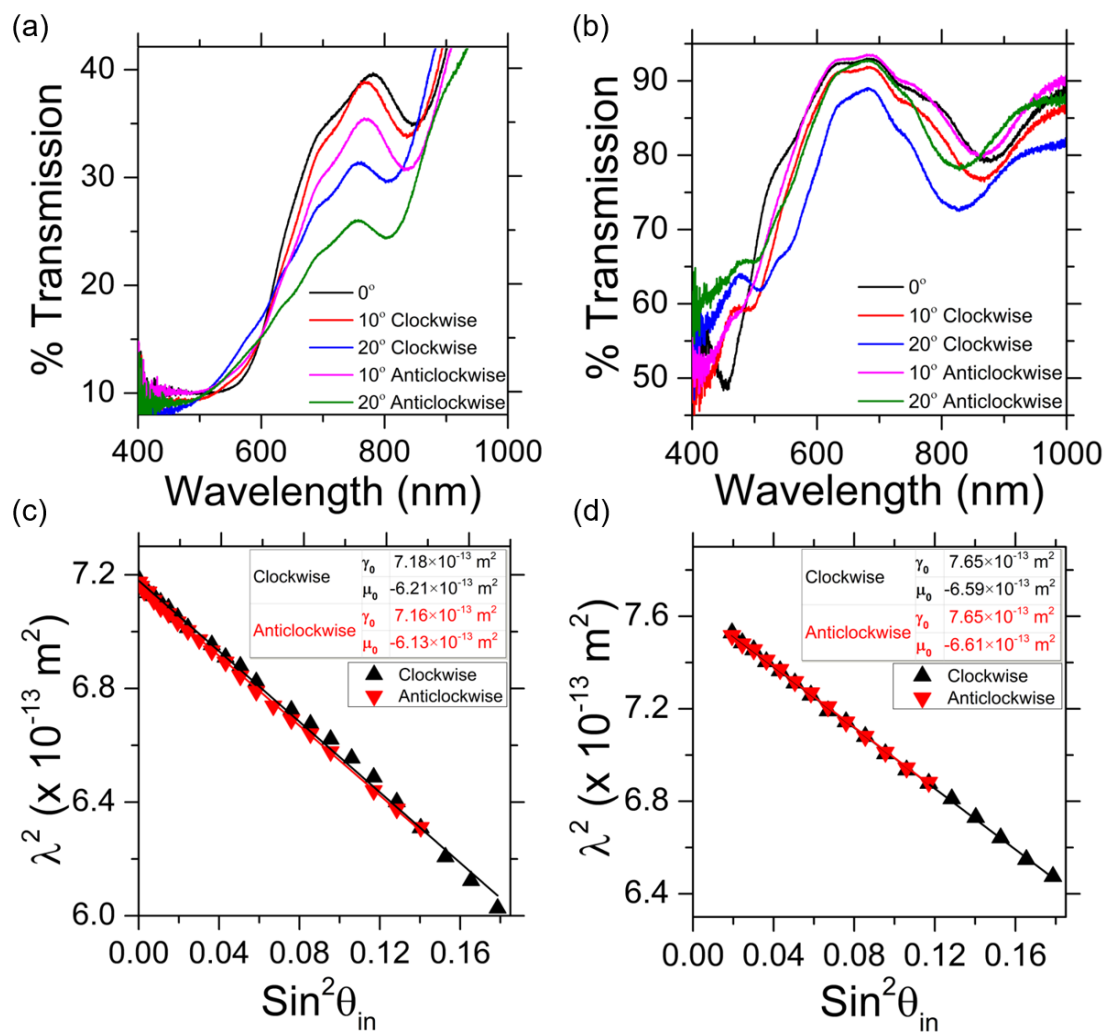

**Figure S6.** Optical transmission spectra for inverse opals immersed in dimethyl carbonate at normal incidence, 10° and 20° (both clockwise and anticlockwise) for a (a) $TiO_2$ and (b) $SnO_2$ IO. Corresponding Bragg-Snell plots of $\lambda_{min}^2$ vs $Sin^2\,\theta_{in}$ in intervals of 1° for both a clockwise and anticlockwise direction about the vertical axis, showing a slope and an intercept value as defined via Eq. (2), for a (c) $TiO_2$ and (d) $SnO_2$ IO.

**Tetrahydrofuran**

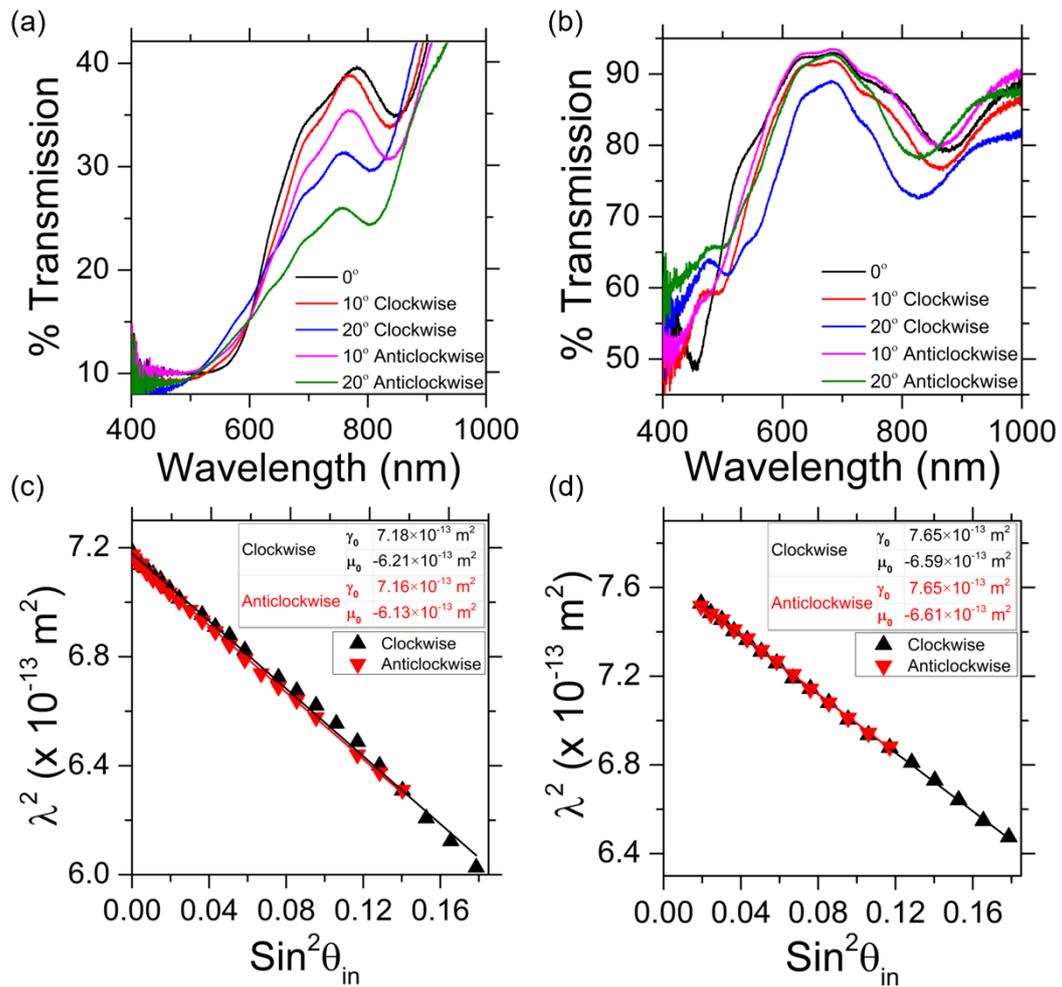

**Figure S7.** Optical transmission spectra for inverse opals immersed in tetrahydrofuran at normal incidence, 10° and 20° (both clockwise and anticlockwise) for a (a) $TiO_2$ and (b) $SnO_2$ IO. Corresponding Bragg-Snell plots of $\lambda_{min}^2$ vs $Sin^2 \theta_{in}$ in intervals of 1° for both a clockwise and anticlockwise direction about the vertical axis, showing a slope and an intercept value as defined via Eq. (2), for a (c) $TiO_2$ and (d) $SnO_2$ IO.

**Dichloromethane**

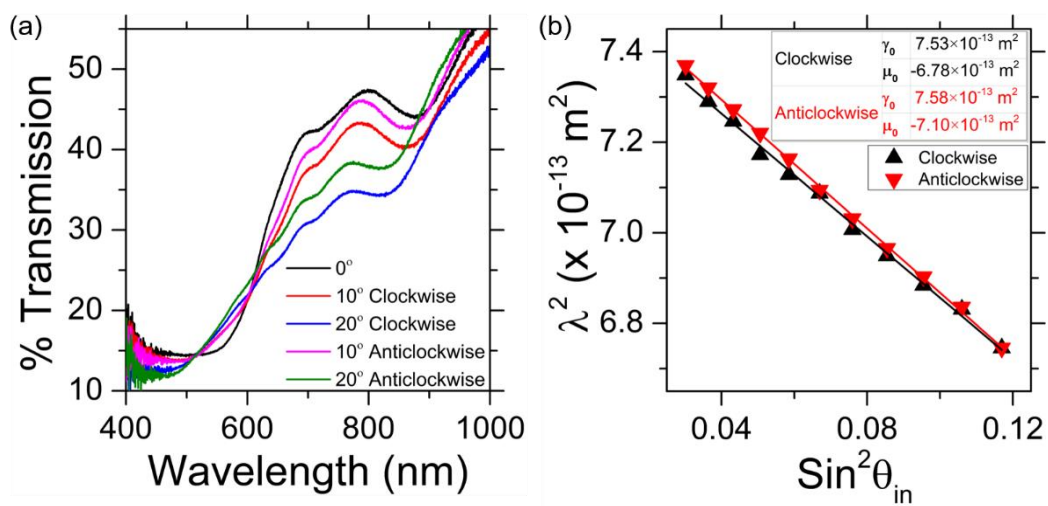

**Figure S8.** (a) Optical transmission spectrum for a TiO$_2$ inverse opal immersed in dichloromethane at normal incidence, 10° and 20° (both clockwise and anticlockwise). (b) Corresponding Bragg-Snell plot of $\lambda_{min}^2$ vs Sin$^2$ $\theta_{in}$ in intervals of 1° for both a clockwise and anticlockwise direction about the vertical axis, showing a slope and an intercept value as defined via Eq. (2), for a TiO$_2$ IO.

**Toluene**

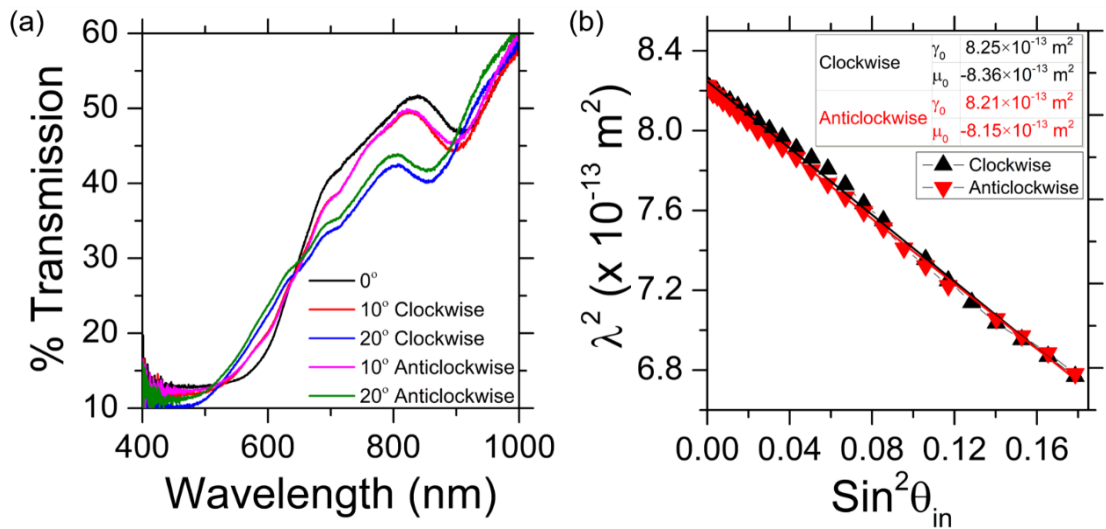

**Figure S9.** (a) Optical transmission spectrum for a TiO$_2$ inverse opal immersed in toluene at normal incidence, 10° and 20° (both clockwise and anticlockwise). (b) Corresponding Bragg-Snell plot of $\lambda_{min}^2$ vs Sin$^2$ $\theta_{in}$ in intervals of 1° for both a clockwise and anticlockwise direction about the vertical axis, showing a slope and an intercept value as defined via Eq. (2), for a TiO$_2$ IO.

**Higher energy transmission minima observed for SnO$_2$ IOs:**

In the main text, a fundamental difference observed between the TiO$_2$ and SnO$_2$ IO optical spectra arises from the interplanar spacing applied to estimate the centre-to-centre pore distances. A reduced spacing of $d = \frac{1}{\sqrt{3}} D$ appeared to give a better estimate of TiO$_2$ pore sizes, compared to the standard spacing of $d = \sqrt{\frac{2}{3}} D$ used to fit the SnO$_2$ IO data. The position of the transmission minima at normal incidence ($\theta = 0°$) were also contrary to expectactions, with TiO$_2$ featuring a minimum at 690 nm and SnO$_2$ featuring a minimum at 720 nm. Using Eq. (2), one would expect the TiO$_2$ IO minimum to be located at longer wavelengths compared to the SnO$_2$ IO. This is due to the larger measured centre-to-centre pore distances for TiO$_2$ (426 nm versus 384 nm) and the larger refractive index of TiO$_2$ (2.488 versus 2.006). However, the opposite proved true in the measured optical spectra with the photonic bandgap of TiO$_2$ being consistently recorded at shorter wavelengths than the SnO$_2$ IO. In an effort to explain this occurrence, the fill factors of the materials were estimated using solvent infilling of the materials at normal incidene and Eq. (9). It was theorised that completely filled materials, such as the SnO$_2$ IO, conform to the standard FCC spacing, whereas less than optimally filled materials, such as the TiO$_2$ IO, adopted a reduced spacing model.

Further optical profiling of the SnO$_2$ IOs revealed a second transmission dip appearing in the optical spectrum in a minority of prepared SnO$_2$ IOs. The appearance of this transmission minimum for the air case can be seen in Fig. S10 (a) at ~605 nm. The transmission data can be seen to be symmetric in the clockwise and anticlockwise directions about normal incidence, effectively removing the possibility of other allowed crystal planes, such as (002) or (220). Other planes would be inclined at a non-zero angle from the normal of the surface and would therefore not feature symmetry about normal incidence. The Bragg-Snell data for this transmission minimum can be seen in Fig. S10 (b), where the data in the clockwise and anticlockwise directions are nearly identical. Additionally, upon analysing the data for the slope and intercept using Eq (12), the reduced spacing model appears to better model the data, providing a centre-to-centre pore esimate of 404 nm compared to 286 nm for the standard spacing model. The effective refractive index can be estimated as 1.294, a reasonable estimate for a SnO$_2$ IO.

The opitical response of the observed second transmission minimum to solvent filling can be seen in Fig. S10 (c). As expected, the position of the transmission minimum shifts to longer wavelengths with increasing solvent refractive index. Just as in the case of the TiO$_2$ and SnO$_2$ IOs in the main text, it was possible to estimate a fill factor for these materials based on this data for both the Parallel and Drude model.

Plots of $\lambda_{min}$ versus $n_{eff}$ for the Parallel and Drude model can be seen in Fig. S10 (d) and (e), respectively. Application of Eq. (9) yields estimates for the volume fill factor of IO material of 16.6% and 10.4% for the Parallel and Drude model, respectively. Compared to optimal fill fraction of 26%, this analysis would suggest a less than optimal filling fraction of IO material for these IO regions, much like in the case of the $TiO_2$ IO. The optical behaviour of the $SnO_2$ IO in these areas is analogous to the $TiO_2$ IO. Both structures appear to feature a less than optimal volume filling fraction with a reduced interplanar spacing required to fit the data. It may be the case that only close to optimally filled IO structures, such as the $SnO_2$ IO with transmission dip at 720 nm, feature the standard interplanar spacing model.

Typically, these transmission minima at lower wavelengths for $SnO_2$ tended to appear closer to the edges of material and exhbitied a different structural colour as seen in Fig. S10 (f). In fact, both recorded minima for the $SnO_2$ IO were recorded from different points on the same IO sample. The transmission minimum at 605 nm was recorded close to the top of the IO sample in Fig. S10 (f) with a yellow/light orange tint. The transmission dip at 720 nm was recorded in the larger central portion of the sample with a dark orange color. SEM images for the 720 nm transmission dip can be seen in Fig. S2. Figures S10 (g) and (f) display the region of the IO associated with the 605 nm transmission dip. Notably, most of this region appears to feature an overlayer of material, making individual centre-to-centre pore distances difficult to assess. The formation of an overlayer is most likely a consequence of excessive precursor build-up on the edges of the polystyrene sphere template[9]. Several layers of material can be seen at the edges of IO domains. It is unclear what effect the overlayer would have on the appearance of the optical spectrum, however, judging from the associated optical data, the spectrum seems to follow the same trends oberserved for the $TiO_2$ IO. During drying, however, overlayers can stretch and curl the edge IO material, and these regions are not taken to be representative of a high quality $SnO_2$ IO in general. The correct structure and analysis are presented in the $SnO_2$ IO data in the main text.

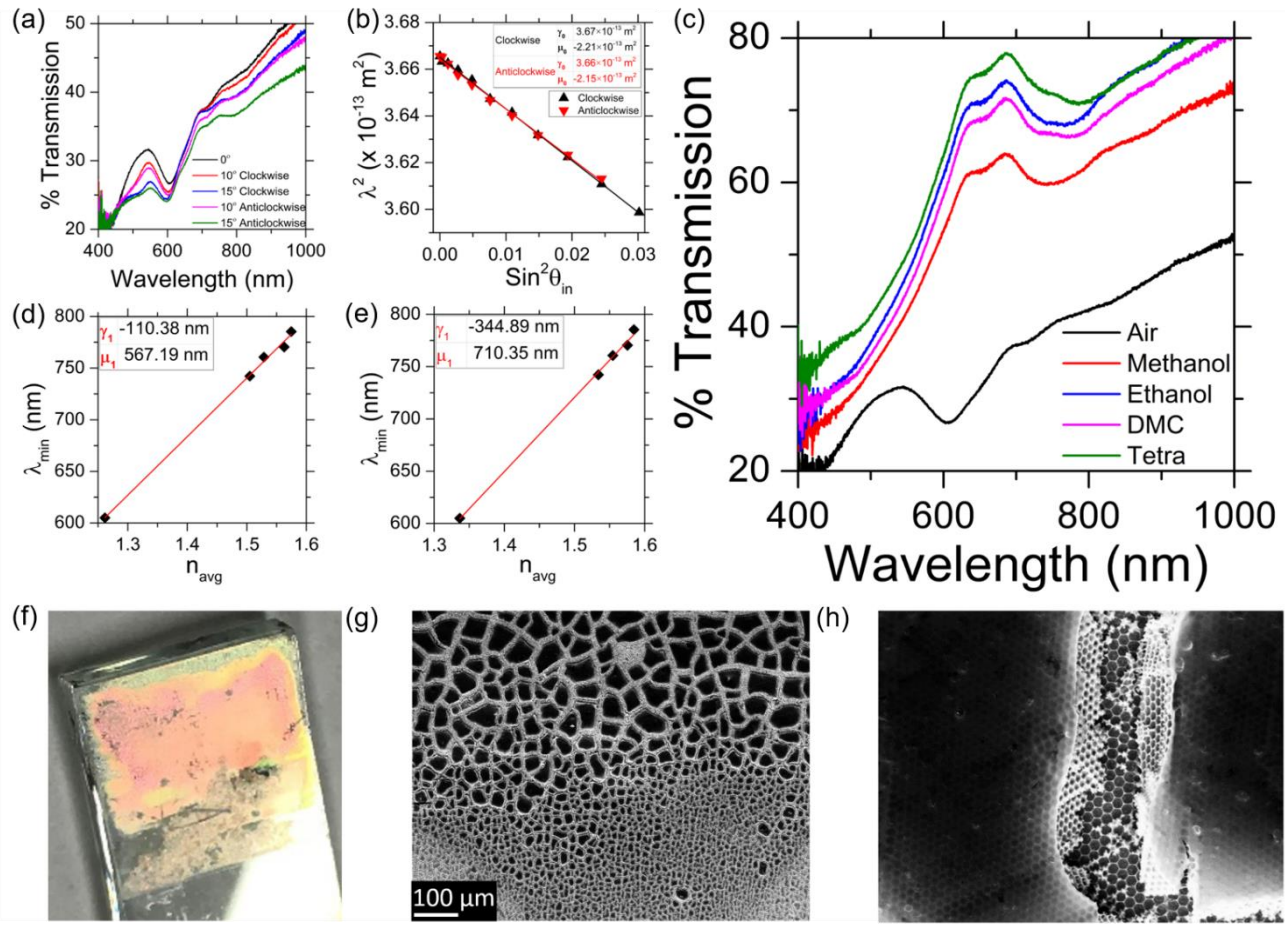

**Figure S10.** (a) Optical transmission spectrum showing a second observed transmission dip for a $SnO_2$ IO. (b) Corresponding Bragg-Snell plot of $\lambda_{min}^2$ vs $Sin^2\theta_{in}$ in intervals of 1° for both a clockwise and anticlockwise direction about the vertical axis. (c) Transmission minima at normal incidence for a $SnO_2$ IO in air, methanol, ethanol, dimethyl carbonate and tetrahydrofuran. Plots of $\lambda_{min}$ versus $n_{eff}$ for a $SnO_2$ using the (d) Parallel and (e) Drude models. (f) optical image of a $SnO_2$ IO showing a variation in reflected colour from the surface. (g) & (h) SEM images showing the area of $SnO_2$ IOs corresponding to a second transmission dip at lower wavelengths.

**Band structure diagrams from modified IO structure simulations:**

Simulated band structure diagrams for custom modelled inverse opal structures computed through the plane wave expansion method using BandSOLVE software are shown through Figs. S11 – S13. Simulations are based on a FCC packed inverse opal with a customisable fill factor of crystalline material, assuming filling of all available voids where the degree of porosity of the material can be varied. Different directions are coloured for ease of viewing, with Γ – M shown in blue, M – K shown in green and K – Γ shown in yellow. Figures S11 (a) and (b) display the band structure diagrams for a $TiO_2$ IO in air with a centre-centre diameter of D = 400 nm for fill factors of $\varphi_{IO} = 0.26$ and $\varphi_{IO} = 0.18$, respectively. In both cases, a full photonic bandgap across all directions is present, however there is a drastic difference in the position and magnitude of the frequency range expected in the photonic bandgap. For $\varphi_{IO} = 0.26$, the simulated photonic bandgap covers a large wavelength range, extending from 735 nm to 844 nm. In terms of the optical transmission spectrum, this would correlate to the appearance of a transmission dip across this range, with the transmission minimum expected in the centre of this wavelength span. In contrast, the simulated photonic bandgap for $\varphi_{IO} = 0.18$ features a noticeably smaller range of forbidden wavelengths extending only from 724 nm to 739 nm. Assuming the centre of the wavelength range spanned by the photonic bandgap acts as a reasonable indication of the position of the transmission minimum, there is a clear blue shift of the transmission minimum in switching from $\varphi_{IO} = 0.26$ to $\varphi_{IO} = 0.18$, going from $\lambda_{min} = 790$ nm to $\lambda_{min} = 731$ nm. In terms of Bragg-Snell analysis, this shift would be incorporated by a change to the average refractive index ($n_{eff}$), by moving to a lower fill fraction of the high index contrast crystalline material. The $SnO_2$ IO was modelled using $\varphi_{IO} = 0.26$ to match both the theoretical maximum and the experimentally determined fill factor. For the $SnO_2$ IO, the photonic bandgap is much narrower than the $TiO_2$ IO case for $\varphi_{IO} = 0.26$, and slightly wider than the $TiO_2$ IO for $\varphi_{IO} = 0.18$. As seen in Fig. S11 (c), the photonic bandgap for the $SnO_2$ IO in air extends from 702 nm to 728 nm, presumably giving a minimum transmission wavelength in the centre of this range at $\lambda_{min} = 715$ nm.

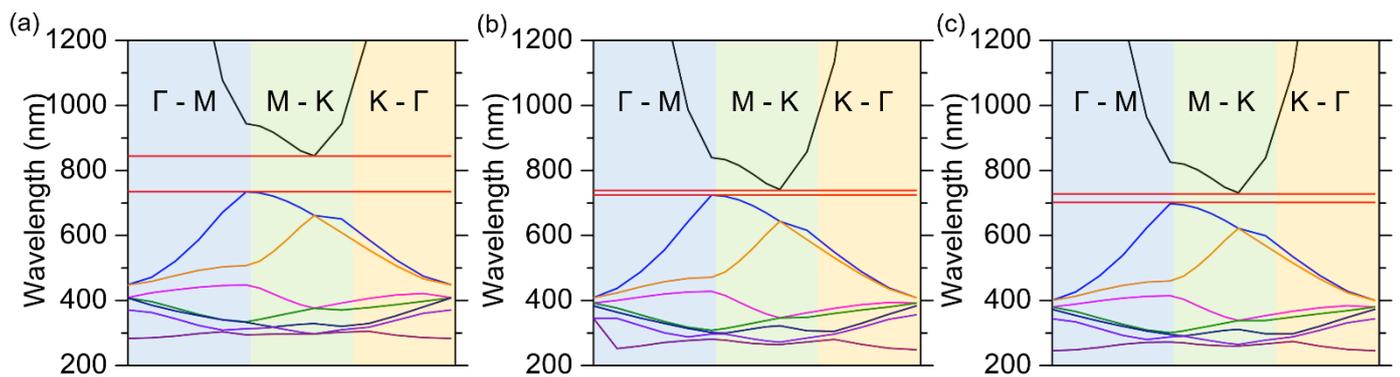

**Figure S11.** Band structure diagrams showing the first 8 bands calculated in air for a D = 400 nm $TiO_2$ IO with customised fill factors of (a) $\varphi_{IO} = 0.26\%$ and (b) $\varphi_{IO} = 0.18\%$. (c) Corresponding band structure diagram for a D = 384 nm $SnO_2$ IO with $\varphi_{IO} = 0.26\%$.

Figure S12 displays the effect of changing the background filling medium on the band structure of a $TiO_2$ IO with $\varphi_{IO} = 0.18$. There is a significant shift in the position of the projected photonic bandgap when changing from air (shown in Fig. S12 (a)) to solvents of higher refractive indices, illustrated here by methanol and toluene in Figs. S12 (b) and (c), respectively. With the addition of a solvent, the band diagrams no longer predict a full photonic bandgap across all directions, instead showing a pseudo-photonic bandgap assigned to the Γ – M direction by convention. The effective refractive index ($n_{eff}$) of the IO is expected to increase when a solvent of higher refractive index replaces the air in structure (see Eqs. (3) and (4)). This is reflected in the transmission spectra via a shift in $\lambda_{min}$ to longer wavelengths. This trend is upheld in the band structure diagrams, where the range of forbidden frequencies extends from 860 – 938 nm for methanol ($n_{sol}$ = 1.329) and from 928 – 991 nm for toluene ($n_{sol}$ = 1.495).

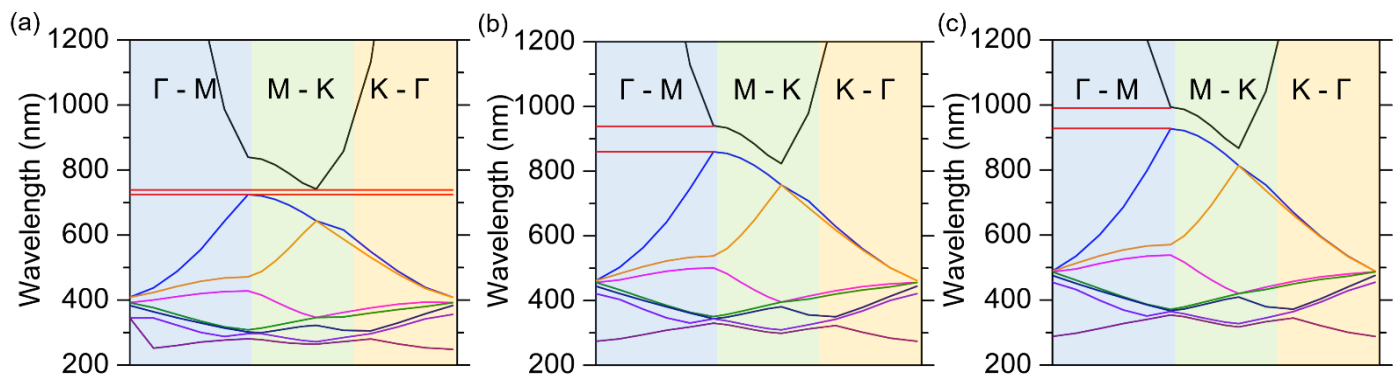

**Figure S12.** Band structure diagrams showing the first 8 bands calculated in a variety of different solvent environments for a D = 400 nm $TiO_2$ IO with φ = 0.18%. Band diagrams are shown calculated for (a) air (n = 1), (b) methanol (n = 1.329) and (c) Toluene (n = 1.495) solvent filling environments.

Figure S13 illustrates the effect of solvent filling on $SnO_2$ IO for $\varphi_{IO}$ = 0.26. In a similar manner to the $TiO_2$ IO with $\varphi_{IO}$ = 0.18, the band structure diagrams for $SnO_2$ no longer predict a full photonic bandgap in the presence of a solvent. The pseudo-photonic bandgap ranges for the $SnO_2$ IO for methanol and toluene are shown in Figs. S13 (b) and (c), respectively. For methanol (n = 1.329), this range can be seen to extend from 856 – 941 nm. In the case of toluene, the wavelength span of the photonic bandgap is projected as 915 – 985 nm. As expected, the wavelength ranges of the simulated photonic bandgaps red-shift with the addition of higher index solvents to the IO medium.

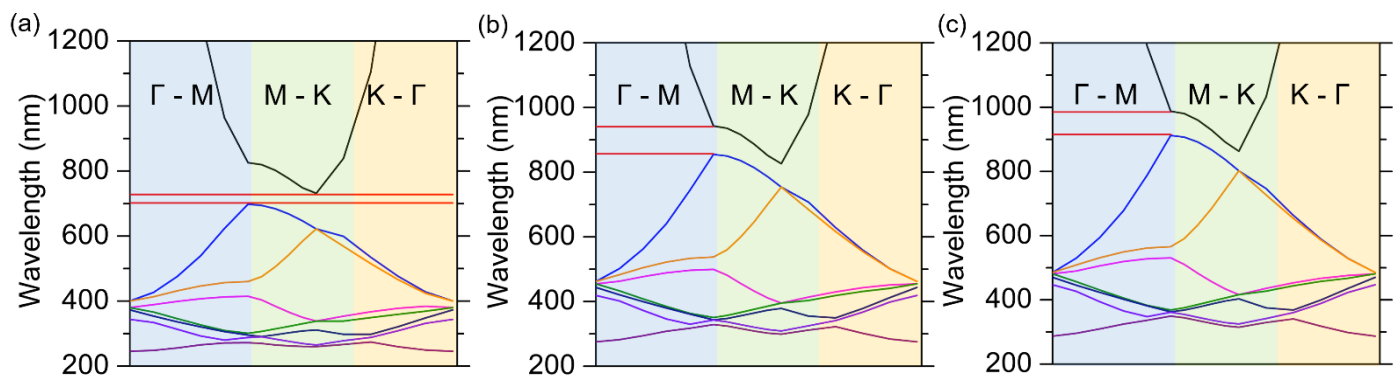

**Figure S13.** Band structure diagrams showing the first 8 bands calculated in a variety of different solvent environments for a D = 384 nm $SnO_2$ IO with $\varphi_{IO}$ = 0.26%. Band diagrams are shown calculated for (a) air (n = 1), (b) methanol (n = 1.329) and (c) Toluene (n = 1.495) solvent filling environments.